\documentclass[onecollarge,final]{svjour2}

\smartqed

\usepackage [dvips]{graphicx}
\usepackage{amssymb}
\usepackage{epsfig}
\usepackage{graphicx}
\usepackage{verbatim}
\usepackage{geometry} 
\usepackage[T1]{fontenc}

\usepackage{amsmath}
\usepackage{enumerate}
\usepackage{graphicx}
\usepackage{mathbbol}
\usepackage{amsfonts}
\usepackage{color}

\def\Xint#1{\mathchoice
{\XXint\displaystyle\textstyle{#1}}%
{\XXint\textstyle\scriptstyle{#1}}%
{\XXint\scriptstyle\scriptscriptstyle{#1}}%
{\XXint\scriptscriptstyle\scriptscriptstyle{#1}}%
\!\int}
\def\XXint#1#2#3{{\setbox0=\hbox{$#1{#2#3}{\int}$}
\vcenter{\hbox{$#2#3$}}\kern-.5\wd0}}

\def\dashint{\Xint-}

\begin{document}


\title{Reunion probability of $N$ vicious walkers: typical and large 
fluctuations for large $N$}


\author{Gr\'egory Schehr \and Satya N. Majumdar \and Alain Comtet \and Peter J. Forrester}

\institute{G. Schehr \at  Laboratoire de Physique Th\'eorique et Mod\`eles
  Statistiques, Universit\'e Paris-Sud, B\^at. 100, 91405 Orsay Cedex, France \\  \and S.~N. Majumdar \at Laboratoire de Physique Th\'eorique et Mod\`eles
  Statistiques, Universit\'e Paris-Sud, B\^at. 100, 91405 Orsay Cedex,
France \\ \and A. Comtet  \at Laboratoire de Physique Th\'eorique et Mod\`eles
  Statistiques, Universit\'e Paris-Sud, B\^at. 100, 91405 Orsay Cedex, France\\
Universit\'e Pierre et Marie Curie-Paris 6, 75005, Paris, France \\ 
\and P. J. Forrester \at Department of Mathematics and Statistics, The University of Melbourne, Victoria 3010, Australia}



\date{\today}

\maketitle

\begin{abstract}

We consider three different models of $N$ non-intersecting Brownian motions on a line segment 
$[0,L]$ with absorbing (model A), periodic (model B) and reflecting (model C) boundary 
conditions. In these three cases we study a properly normalized reunion probability, which, 
in model A, can also be interpreted as the maximal height of $N$ non-intersecting Brownian 
excursions (called "watermelons" with a wall) on the unit time interval. We provide a 
detailed derivation of the exact formula for these reunion probabilities for finite $N$ 
using a Fermionic path integral technique. We then analyse the asymptotic
behavior of this reunion probability for large $N$ using two
complementary techniques: (i) a saddle point analysis of the underlying Coulomb gas and
(ii) orthogonal polynomial method. These two methods are complementary in the sense
that they work in two different regimes, respectively for $L\ll O(\sqrt{N})$
and $L\geq O(\sqrt{N})$. 
A striking feature of 
the large $N$ limit of the reunion probability in the three models is that it exhibits a 
third-order phase transition when the system size $L$ crosses a critical value $L=L_c(N)\sim 
\sqrt{N}$. This transition is akin to the Douglas-Kazakov transition in two-dimensional 
continuum Yang-Mills theory.  While the central part of the reunion probability, for $L \sim 
L_c(N)$, is described in terms of the Tracy-Widom distributions (associated to GOE and GUE 
depending on the model), the emphasis of the present study is on the large deviations of 
these reunion probabilities, both in the right [$L \gg L_c(N)$] and the left [$L \ll L_c(N)$] 
tails. In particular, for model B, we find that the matching between the different regimes 
corresponding to typical $L \sim L_c(N)$ and atypical fluctuations in the right tail $L \gg 
L_c(N)$ is rather unconventional, compared to the usual behavior found for the distribution 
of the largest eigenvalue of GUE random matrices. This paper is an extended version of [G. 
Schehr, S. N. Majumdar, A. Comtet, J. Randon-Furling, Phys. Rev. Lett. {\bf 101}, 150601 
(2008)] and [P.~J.~Forrester, S.~N.~Majumdar, G.~ Schehr, Nucl. Phys. B {\bf 844}, 500-526 
(2011)].

\end{abstract}

\section{Introduction}

Non-intersecting random walkers, first introduced
by de Gennes~\cite{deG68}, followed by Fisher~\cite{Fisher84} (who called them "vicious walkers"), 
have been studied extensively
in statistical physics as they appear in a variety of physical contexts 
ranging from wetting and melting all the way to polymers and vortex lines
in superconductors. Lattice versions of such walkers also have beautiful combinatorial properties~\cite{KGV2000}.
Non-intersecting Brownian motions, defined in continuous space
and time, have also recently appeared in a number of contexts. In
particular their connection to the random matrix theory
has been noted in a variety of situations~\cite{Ba00,Fo01,Jo03,Na03,KT2004,FP2006,TW2007,DaKu07,SMCR08,NM09,No09,RS10,BoKu10,BDK11,AMV12}. 
Rather recently, we have unveiled an unexpected connection between
vicious walkers and two-dimensional continuum Yang-Mills theory 
on the sphere with a given gauge group $G$ \cite{FMS11}, 
which depends on the boundary conditions in the vicious walkers 
problem (a different type of connection between the Yang-Mills theory and vicious walkers 
problems had also been noticed in 
Ref. \cite{HT04}).  

Specifically, we consider a set of $N$ non-intersecting Brownian 
motions on a finite segment~$[0,L]$ of the real line with
different boundary conditions.
Assuming that all the walkers start from the vicinity of 
the origin, we then
define the reunion probability as the probability that the walkers
reunite at the origin after a fixed interval of time which 
can be set to unity without any loss of generality.
Within the time interval $[0,1]$, the walkers stay non-intersecting. 
Next we `normalize' this reunion probability in a precise way to
be defined shortly. 
In one case, namely when both boundaries at $0$ and $L$ are absorbing,
one can relate this `normalized' reunion probability to the probability 
distribution
of the maximal height of $N$ non-intersecting Brownian excursions. In Ref. \cite{FMS11} it was shown 
that this normalized reunion probability
in the Brownian motion models maps onto the exactly solvable partition function (up to a
multiplicative factor) of two-dimensional 
Yang-Mills theory on a sphere. The boundary conditions at the edges $0$
and $L$ select the gauge group $G$ of the associated Yang-Mills theory.
We consider three different boundary conditions: absorbing (model A), periodic (model B), 
 and reflecting (model C) which correspond respectively
to the following gauge groups in the Yang-Mills theory: (A) absorbing $\to$ 
${\rm Sp}(2N)$, (B) periodic $\to$ 
${\rm U}(N)$  and (C) reflecting $\to$ ${\rm SO}(2N)$. As a consequence of this connection, in each of these Brownian motion 
models, as one varies the system size $L$, a third order phase transition occurs at a critical value $L = L_c(N) = {\cal O}(\sqrt{N})$ in the large $N$
limit. It was shown in Ref.~\cite{NM11} that a similar third order phase transition also 
occurs for the probability distribution 
function of the largest eigenvalue of random matrices belonging to the 
standard Gaussian ensembles. Furthermore, third order phase transitions 
in the large deviation function of appropriate variables have also been found recently
in a variety of other problems, such as in the distribution of conductance
through a mesoscopic cavity such as a quantum dot~\cite{VMB08,VMB10,DMTV11} and in
the distribution of the
entanglement 
entropy
in a bipartite random pure state~\cite{NMV10,NMV11}.

Close to the critical point, these reunion probabilities in the Brownian motion models, 
properly shifted and scaled, 
can be related to the
Tracy-Widom (TW) distributions.
Let us briefly remind the readers about the Tracy-Widom distributions.
Tracy-Widom distribution describes the limiting form of the
distribution 
of the scaled largest eigenvalue in 
the three classical Gaussian random matrix ensembles~\cite{TW94a}. 
These limiting distributions are usually denoted by 
$\mathcal{F}_1(t)$ for the Gaussian orthogonal ensemble (GOE),
by $\mathcal{F}_2(t)$ for the Gaussian unitary ensemble (GUE) and
by $\mathcal{F}_4(t)$ for the Gaussian symplectic ensemble (GSE). 
For example, in the GUE case where one considers the set
of $N \times N$ complex Hermitian matrices $X$ with measure 
proportional to $e^{-{\rm Tr} X^2}$ and denote $\lambda_{\max}$
the largest eigenvalue, the typical fluctuations around its 
mean value $\langle \lambda_{\max} \rangle \simeq \sqrt{2N}$ 
have a limiting distribution \cite{TW94a}
\begin{equation} \label{F2}
\lim_{N\rightarrow \infty} {\rm Pr}\bigg({\sqrt{2}N^{\frac{1}{6}}} (\lambda_{\rm 
max}-\sqrt{2N})<t \bigg)
=\exp\bigg( - \int_t^\infty (s-t)q^2(s)ds \bigg):= \mathcal{F}_2(t),
\end{equation}
known as the $\beta=2$ Tracy-Widom distribution. 
In Eq. (\ref{F2}), $q(s)$ satisfies Painlev\'e II (PII) differential equation
\begin{eqnarray}\label{PII}
q''(s) = s q(s) + 2 q^3(s) \;,
\end{eqnarray}
with the asymptotic behavior
\begin{eqnarray}\label{asympt_q}
q(s) \sim {\rm Ai}(s)\, {\rm as}\,\, s\to \infty \;,
\end{eqnarray}
where ${\rm Ai}(s)$ is the Airy function. 
One can indeed show that this asymptotic behavior 
(\ref{asympt_q}) determines a unique solution of PII (\ref{PII}), 
known as the Hastings-McLeod solution. Similarly, 
the distribution $\mathcal{F}_1$ concerns real symmetric matrices. Explicitly, 
with the GOE specified as the set of $N \times N$ real symmetric matrices $X$ with measure 
proportional to $e^{-{\rm Tr }X^2/2}$, and $\lambda_{\rm max}$ denoting the 
largest eigenvalue, one has \cite{TW96}
\begin{eqnarray}\label{F2-GOE}
\lim_{N\rightarrow \infty} 
{\rm Pr} \bigg(\sqrt{2}N^{\frac{1}{6}}(\lambda_{\rm max}-\sqrt{2N})<t\bigg)&=&
\exp\bigg( -\frac{1}{2} \int_t^\infty \left( 
\left(s-t\right)q^2(s)+q(s)\right)\,ds 
\bigg) \nonumber \\ 
&\mathrel{\mathop:}=& \mathcal{F}_1(t) \;,
\end{eqnarray}
known as the $\beta=1$ Tracy-Widom distribution. Because of their relevance in many fundamental problems in mathematics and physics, these TW distributions have been widely studied in the literature. In particular, their asymptotic behaviors are given, to leading order by
\begin{eqnarray}\label{asympt_behaviors}
\begin{cases}
{\cal F}_\beta(t) \sim \exp{\left(-\frac{\beta}{24} |t|^3\right)} \;, \; t \to -\infty \;,\\
& \\
1-{\cal F}_\beta(t) \sim \exp{\left(- \frac{2 \beta}{3} t^{3/2}\right)} \;, \; t \to \infty 
\end{cases}
\end{eqnarray}
where $\beta=1$ and $\beta=2$ correspond respectively to the GOE and the GUE case.

Our study of constrained vicious walkers problems started in Ref. \cite{SMCR08} where we 
derived, using a Fermionic path integral method, an exact 
expression for the ratio of reunion probability, for model A and for any 
finite number $N$ of walkers. This calculation, in this specific case, was motivated by the 
interpretation of this ratio in terms of an extreme value quantity, namely the maximal height 
of $N$ non-intersecting Brownian excursions, which is recalled below in section 
\ref{sec:models} (see also Fig. \ref{fig:watermelon}). We showed later in Ref. \cite{FMS11} 
that this ratio, and its extension to other boundary conditions in model B and C, is actually 
equal, up to a multiplicative prefactor, to the partition function of Yang-Mills theory on 
the sphere with a given gauge group $G$, which depends on the boundary conditions in the 
vicious walkers problem (see Table \ref{table}). Following the pioneering works of Refs. \cite{DK93,GM94} where the 
large $N$ analysis of Yang-Mills theory on the sphere with the gauge group $G={\rm U}(N)$, 
and extended to other classical Lie groups, including $G={\rm Sp}(2N)$ or $G={\rm SO}(2N)$, 
in Ref. \cite{CNS96}, we could also perform in Ref.~\cite{FMS11} the large $N$ analysis of 
these reunion probabilities. However, most of these results were announced in 
Refs.~\cite{SMCR08,FMS11} without any detail. The purpose of the present paper is to 
give a self-contained derivation of these results, both for finite and large $N$. For large 
$N$ we characterize not only the distribution of typical fluctuations, which can be expressed 
in terms of ${\cal F}_1$ and ${\cal F}_2$ but also provide a detailed analysis of the large 
deviations of these reunion probabilities, characterizing atypical fluctuations. A special 
emphasis is put on
the matching between different regimes (typical and atypical) of the reunion probability
as a function of $L$. In particular, for the case of periodic 
boundary conditions, we find that the matching found in this vicious walker problem is quite 
different from the corresponding matching observed in the distribution of the largest 
eigenvalue of Gaussian random matrices in the unitary 
ensemble~\cite{NM09,DM06,VMB07,DM08,MV09,BEMN10,Fo12}.

The paper is organized as follows. In section 2 we describe the three models A, B and C and 
summarize  the main results for the ratio of reunion probabilities in the large $N$ limit. In 
section 3 we give the details of the path integral method which allows us to obtain an exact 
expression of these ratios in each of the three models for any finite $N$ and $L$. In section 
4, we compute the large 
$N$ limit of this ratio for model A and analyze in detail the typical fluctuations (namely 
the central part of the distribution) as well as the large deviations: both for the left and 
for the right tails. The corresponding large $N$ analysis for model B and C are performed in 
section 5 and section 6, respectively, before we conclude in section 7. Some details have 
been relegated to Appendix A.

\section{Models and main results}\label{sec:models}

We consider three different models of $N$ non-intersecting Brownian walkers on a 
one-dimensional line segment $[0,L]$ and label their positions at time $\tau$ by 
$x_1(\tau)<\ldots<x_N(\tau)$. These three models, denoted by model A, B and C, differ by the 
boundary conditions which are imposed at $x=0$ and $x=L$: \\ $\bullet$ in model A, we 
consider absorbing boundary conditions both at $x=0$ and $x=L$, \\ $\bullet$ in model B, we 
study periodic boundary conditions, which amounts to consider $N$ non-intersecting Brownian 
motions on a circle of radius $L/2 \pi$,\\ $\bullet$ and in model C we consider reflecting 
boundary conditions both at $x=0$ and $x=L$.

\vskip 0.3cm

\begin{table}[h]
\begin{center}
\label{fig_table}
\begin{tabular}{|c||c|c|c|}
\hline
\quad & \quad & \quad & \quad \\
 \quad & Model A & Model B & Model C \\
 \quad & \quad & \quad & \quad \\
\hline
\hline
\quad & \quad & \quad & \quad \\
Boundary conditions at $x=0$ and $x=L$ & absorbing & periodic & reflecting \\
\quad & \quad & \quad & \quad \tabularnewline
\hline
\quad & \quad & \quad & \quad \\
Corresponding affine Weyl chamber & $\tilde C_N$ & $\tilde A_{N-1}$ & $\tilde B_{N}$ \\
\quad & \quad & \quad & \quad \\
\hline
\quad & \quad & \quad & \quad \\
Gauge group of the associated YM$_2$ theory& ${\rm Sp}(2N)$ & ${\rm U}(N)$ & ${\rm SO}(2N)$ \\
\quad & \quad & \quad & \quad \\
\hline
\end{tabular}
\end{center}
\hspace*{2cm} \caption{Summary of our results for the three different models A, B and C.}\label{table}
\end{table}

\noindent {\bf Model A:} In the first model the domain is the 
line segment $[0,L]$ with {\em absorbing boundary conditions at both
boundaries $0$ and $L$}. This corresponds to $N$-dimensional Brownian motion
in an affine Weyl chamber of type $\tilde C_N$~\cite{Gra99,Gra02}. The $N$ non-intersecting Brownian
motions start initially at the positions, say, $\{\epsilon_1, 
\epsilon_2,\ldots, \epsilon_N\}$ in the vicinity of the origin where
eventually we will take the limit $\epsilon_i\to 0$ for all $i$, as shown later. 
We define the reunion probability 
$R_L^{A}(1)$, where the superscript 'A' refers to model A, as
the probability that the walkers return to their initial positions
after a fixed time $\tau=1$ (staying non-intersecting over the time interval
$\tau\in [0,1]$). We define the normalized reunion probability
\begin{equation}\label{RL}
\tilde{F}_N(L)={R_L^{A}(1) \over R_\infty^{A}(1)} \;,
\end{equation}
such that $\lim_{L \to \infty} \tilde F_N(L) = 1$. This ratio
becomes independent of the starting positions $\epsilon_i$'s
in the limit when $\epsilon_i\to 0$ for all $i$, as shown later.
Hence, $\tilde{F}_N(L)$ depends only on $N$ and $L$.  

\begin{figure}
\begin{center}
\includegraphics[width=0.7\linewidth]{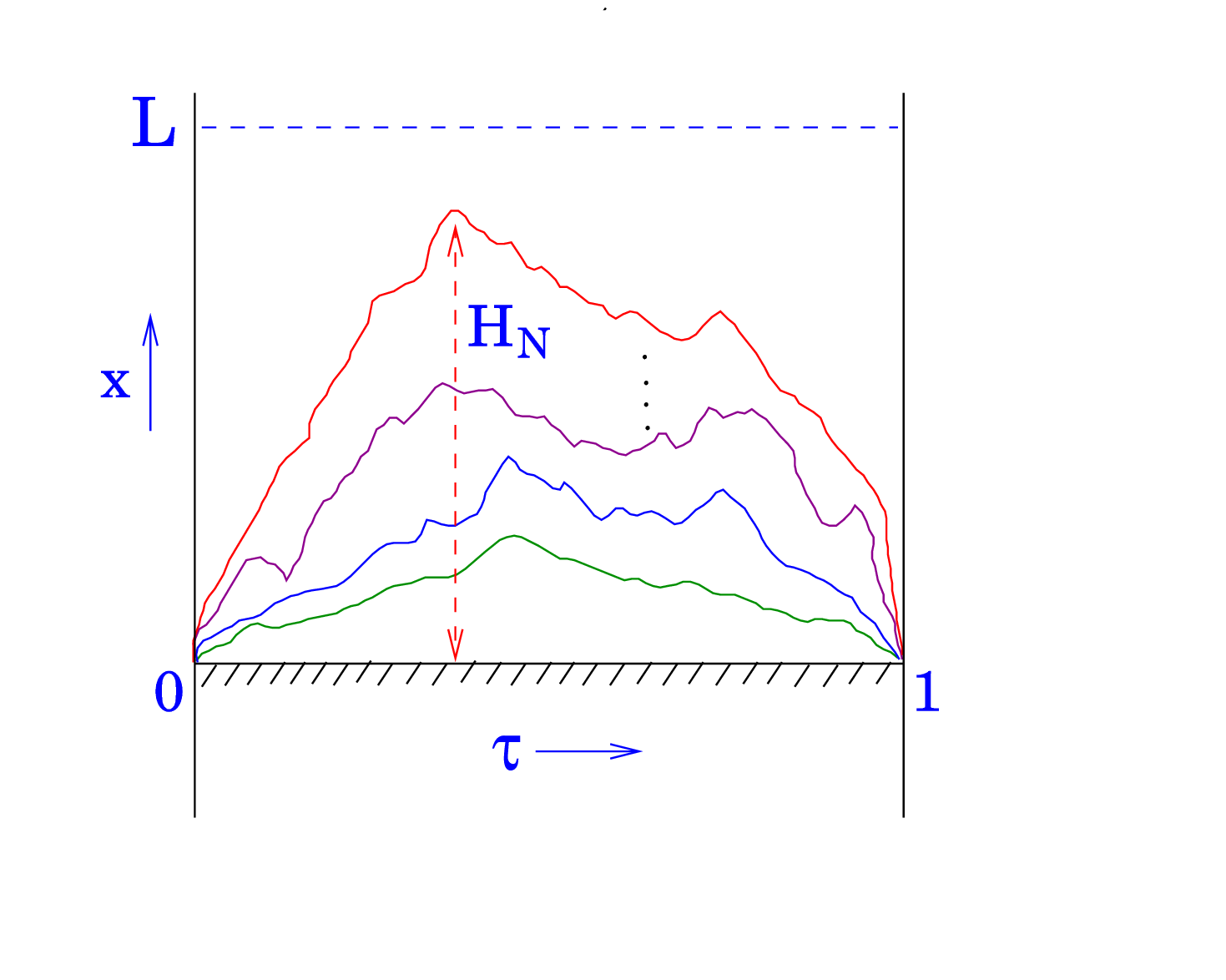}
\caption{Trajectories of $N$ non-intersecting Brownian motions 
$x_1(\tau)<x_2(\tau)<\ldots < x_N(\tau)$, all start at the origin
and return to the origin at $\tau=1$, staying positive in between.
$\tilde{F}_N(L)$ denotes the probability
that the maximal height $H_N=\max_\tau\{x_N(\tau), 0\le \tau\le 1\}$ stays below 
the level $L$ 
over the time interval $0\le \tau\le 1$.}
\label{fig:watermelon}
\end{center}
\end{figure}

This ratio $\tilde{F}_N(L)$ in Model A has also a different 
probabilistic interpretation. Consider the same model but now on
the semi-infinite line $[0,\infty)$ with still absorbing boundary condition 
at $0$. The walkers, as usual, start in the vicinity of the origin 
and are conditioned to return to the origin exactly at $\tau=1$ (see Fig.~\ref{fig:watermelon}). 
If one plots the space-time trajectories of the walkers, a typical
configuration looks like half of a watermelon
(see Fig. \ref{fig:watermelon}), or a watermelon in presence
of an absorbing wall. Such configurations of Brownian motions
are known as non-intersecting Brownian excursions~\cite{TW2007} and their
statistical properties have been studied quite extensively in
the recent past. A particular observable that has  
generated some recent interests is the so-called `height'
of the watermelon~\cite{SMCR08,BM2003,Fulmek2007,KIK2008,KIK08,Fe08,RS11,liechty} defined
as follows (see also Ref. \cite{BFPSW09} for a related quantity in the 
context of Dyson's Brownian motion). Let $H_N$ denote the 
maximal displacement of the rightmost walker $x_N$ in this time interval 
$\tau\in [0,1]$, i.e., the maximal height of the topmost path
in half-watermelon configuration (see Fig. \ref{fig:watermelon}), i.e.,
$H_N=\max_\tau\{x_N(\tau),
0<\tau<1\}$. 
This global maximal height $H_N$   
is 
a random variable which fluctuates from one configuration of half-watermelon to 
another. What is the probability distribution of $H_N$? For $N=1$ the 
distribution of $H_N$ is easy to compute and already for $N=2$ it is somewhat
complicated~\cite{KIK2008}. In Ref.~\cite{SMCR08} an exact formula for the 
distribution of $H_N$, valid for all $N$, was derived using Fermionic path integral method, which we remind below for consistency (see also \cite{KIK2008,KIK08,Fe08} for a derivation using different methods). The distribution of $H_N$, in the large $N$ limit, is quite interesting as it gives, in the proper scaling limit, the distribution of the maximum (on the real line) of the Airy$_2$ process minus a parabola \cite{Jo03,PS99,PS04}. This latter process describes the universality class of the Kardar-Parisi-Zhang (KPZ) equation in the so-called "droplet" geometry. It was known rather indirectly from the work of Ref. \cite{Jo03} on the Airy$_2$ process
that the limiting distribution of $H_N$ should then be given by ${\cal F}_1$. 
This was one of the main result of Ref. \cite{FMS11} to show this result by a 
direct computation of the limiting distribution of $H_N$. 
This result, for the vicious walkers problem, was then recently proved rigorously in 
\cite{liechty} using Riemann-Hilbert techniques. In Ref.~\cite{MQR11} a direct and rigorous 
proof was established that the distribution of the maximum of Airy$_2$ minus a parabola is 
indeed given by ${\cal F}_1$. We refer the reader to Refs. \cite{Sc12,QR12,BLS12} for more 
recent results on the extreme statistics of the Airy$_2$ process minus a parabola. We also 
mention that the limiting distribution of $H_N$ was measured in recent experiments on liquid 
crystals, and a very good agreement with ${\cal F}_1$ was indeed found \cite{TS12}.

\begin{figure}[ht]
\begin{center}
 \includegraphics[width = \linewidth]{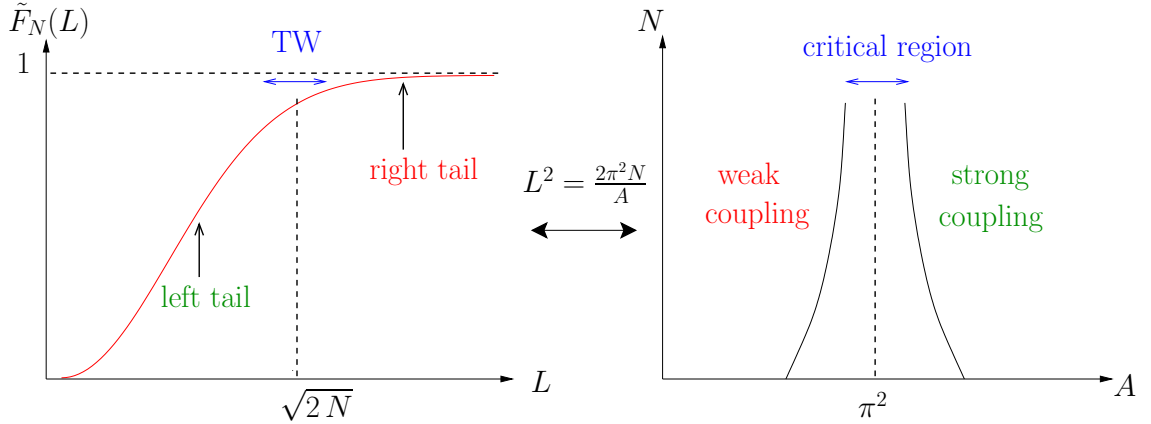}
\caption{{\bf Left:} 
Schematic sketch of $\tilde F_N(L)$ as defined in Eq. (\ref{Fp}) for $N$ 
vicious 
walkers on the line segment $[0,L]$ with absorbing boundary conditions at both ends, 
as a function of $L$, for fixed but large $N$.  {\bf 
Right:} Sketch of the phase 
diagram in the plane 
$(A,N)$ of two-dimensional Yang-Mills theory on a sphere with the gauge 
group ${\rm Sp}(2N)$ as obtained in Ref. \cite{DK93,CNS96}. The weak (strong)
coupling phase in the right panel corresponds to the right (left)
tail of $\tilde F_N(L)$ in the left panel. The critical region around
$A=\pi^2$ in the right panel corresponds to the Tracy-Widom (TW) regime
in the left panel around the critical point $L_c(N)= \sqrt{2N}$.}   
\label{fig_corres}
\end{center}
\end{figure}

To relate the distribution of $H_N$ in the semi-infinite system
defined above to the ratio of
reunion probabilities in the finite segment $[0,L]$ defined in
\eqref{RL}, it is  
useful to consider the cumulative probability ${\rm Pr}(H_N\leq L)$ 
in the semi-infinite geometry, where
$L$ now is just a variable. To compute this cumulative probability, we need
to calculate the fraction of half-watermelon configurations (out of all
possible half-watermelon configurations) that 
never
cross the level $L$, i.e., whose heights stay below $L$ over the time interval 
$\tau\in [0,1]$ (see Fig. \ref{fig:watermelon}). This fraction can be 
computed by putting an absorbing boundary at $L$ (thus killing all 
configurations that touch/cross the level $L$). It is then clear
that ${\rm Pr}(H_N\leq L)$ is nothing but the normalized reunion probability
${\tilde F}_N(L)$ defined in \eqref{RL}. As mentioned above, this 
cumulative probability 
distribution of the maximal height was computed exactly in Ref. \cite{SMCR08}
\begin{align}\label{Fp}
\tilde{F}_N(L)&:= {\rm Pr}(H_N\leq L) \nonumber \\
&=\frac{A_N}{L^{2N^2+N}}\sum_{n_1=-\infty}^\infty \ldots \sum_{n_N=-\infty}^{\infty} 
\Delta^2(n_1^2,\ldots,n_N^2)\Big(\prod_{j=1}^N n_j^2\Big)
e^{-\frac{\pi^2}{2L^2}\sum_{j=1}^N 
n_j^2} \;,
\end{align}
where
\begin{equation}
\Delta_N(y_1,\ldots, y_N)=\prod_{1\leq i<j\leq N}(y_i-y_j) 
\label{vdm}
\end{equation}
is the Vandermonde determinant and where the amplitude is given by
\begin{eqnarray}\label{BN_intro}
A_N=\frac{\pi^{2N^2+N}}{2^{N^2+N/2}\prod_{j=0}^{N-1}\Gamma(2+j)\Gamma(3/2+j)} \;.
\end{eqnarray}
This quantity $\tilde F_N(L)$ was also computed in Refs.~\cite{KIK08,Fe08} using different methods. In Ref.~\cite{KIK08} the authors 
studied the maximal height of $N$ non-intersecting Brownian excursions and used the Karlin-McGregor formula \cite{karlin_mcgregor}. In Ref. \cite{Fe08} the author computed the cumulative distribution of the maximal height of $N$-non intersecting discrete lattice paths (of $n$ discrete steps) in presence of a wall, using the Lindstr\"om-Gessel-Viennot formula \cite{LGV1,LGV2}, and then he considered the asymptotic limit of a large number of steps $n$. The formulas obtained using these two similar methods are actually different from the one given in Eq. (\ref{Fp}) obtained by using a fermionic path integral and it is a non-trivial task to check, as expected, that they are indeed equivalent to (\ref{Fp})~\cite{KIK08}. A derivation of the formula given above in Eqs. (\ref{Fp}, \ref{BN_intro}), which turns out to be the most convenient expression of $\tilde F_N(L)$ in view of a large $N$ asymptotic analysis, is given in section \ref{derivation}.
Remarkably, if one denotes~by 
\begin{eqnarray}\label{Z_YM}
{\cal Z}(A,G) 
\end{eqnarray}
the partition function of the two-dimensional (continuum) 
Yang-Mills theory on the sphere (denoted as $YM_2$) with gauge group 
$G$ and area $A$ it was shown in Ref. \cite{FMS11} that 
$\tilde F_N(L)$ is related to $YM_2$ with the gauge group $G = {\rm Sp}(2N)$ via the relation
\begin{eqnarray}\label{relation_A_YM}
\tilde F_{N}(L) \propto {\cal Z}\left(A = \frac{2 \pi^2}{L^2}\, N, {\rm Sp}(2N) \right) \;.
\end{eqnarray}
In Ref. \cite{DK93,CNS96}, it was shown that for large $N$, 
${\cal Z}(A,{\rm Sp}(2N))$ exhibits a third order phase transition at the 
critical value $A = \pi^2$ separating a weak coupling regime for 
$A < \pi^2$ and a strong coupling regime for $A > \pi^2$. 
This is the so called Douglas-Kazakov phase transition \cite{DK93}, 
which is the counterpart in continuum space-time, of the Gross-Witten-Wadia 
transition~\cite{GW80,Wadia80} found in two-dimensional lattice quantum 
chromo-dynamics (QCD), which is also of third order. 
Using the correspondence $L^2 = 2 \pi^2N/A$, we then find that $\tilde F_N(L)$, 
considered as a function of $L$ with $N$ large but fixed, also exhibits a third order 
phase transition at the critical value $L_c(N) = \sqrt{2N}$. 
Furthermore, the weak coupling regime ($A < \pi^2$) corresponds to $L > \sqrt{2N}$ and thus 
describes the right tail of $\tilde F_N(L)$, while the strong coupling regime corresponds to 
$L < \sqrt{2N}$ and describes instead the left tail of $\tilde F_N(L)$ (see Fig. 
\ref{fig_corres}). The critical regime around $A = \pi^2$ is the so called "double scaling" 
limit in the matrix model and has width of order $N^{-2/3}$. It corresponds to the region of 
width ${\cal O}(N^{-1/6})$ around $L = \sqrt{2N}$ where $\tilde F_N(L)$, correctly shifted 
and scaled, is described by the Tracy-Widom distribution~${\cal F}_1(t)$ in
Eq. (\ref{F2-GOE}).

Although the occurrence of the Painlev\'e transcendent $q(s)$ (\ref{PII}) in this double 
scaling limit was known since the work of Periwal and Shevitz \cite{PS90a}, its probabilistic 
interpretation in relation to the Tracy-Widom distribution was one of the main 
achievements of 
Ref. \cite{FMS11}. In this paper, we provide a detailed analysis of the three regimes: the 
left tail, the central part and the right tail of $\tilde F_N(L)$. Our results can be 
summarized as 
follows
\begin{eqnarray}\label{main_results_F}
\begin{cases}
\tilde F_N(L) &\sim \exp{\left[-N^2 \phi_-^A\left(L/\sqrt{2N} \right) \right]} \;, \; L < \sqrt{2N} \; \& \; |L- \sqrt{2N}| \sim {\cal O}(\sqrt{N}) \\
& \\
\tilde F_N(L) &\sim {\cal F}_1\left[2^{11/6} N^{1/6} (L-\sqrt{2N}) \right] \;, \; L \sim \sqrt{2N} \; \& \; |L-\sqrt{2N}| \sim {\cal O}(N^{-1/6}) \\
& \\
1 - \tilde F_N(L) &\sim \exp{\left[-N \phi_+^A\left(L/\sqrt{2N} \right) \right]} \;, \; L > \sqrt{2N} \; \& \; |L- \sqrt{2N}| \sim {\cal O}(\sqrt{N}) \;,\\
\end{cases}
\end{eqnarray}
where ${\cal F}_1$ is the TW distribution for GOE, whose explicit expression 
is given in (\ref{F2-GOE}) and its asymptotic behaviors are given 
in Eq. (\ref{asympt_behaviors}). The rate functions $\phi_\pm^A(x)$ can be 
computed exactly (see later): of particular interest are their 
asymptotic behaviors when $L \to \sqrt{2 N}$ from below 
(left tail) and from above (right tail), which are given by
\begin{eqnarray}\label{asympt_rate}
\phi_-^A(x) &\sim& \frac{16}{3}(1-x)^3 \;, \; x \to 1^- \;, \\
\phi_+^A(x) &\sim& \frac{2^{9/2}}{3}(x-1)^{3/2}\;, \; x \to 1^+ \;. \nonumber
\end{eqnarray}
The different behavior of $\tilde F_N(L)$ in Eq. (\ref{main_results_F}) 
for $L < \sqrt{2N}$ and $L > \sqrt{2N}$ leads, in the limit $N \to \infty$ to a 
phase transition at the critical point $L = \sqrt{2N}$ in the following sense. 
Indeed if one scales $L$ by $\sqrt{2N}$, keeping the ratio $x = L/\sqrt{2N}$ fixed, 
and take the limit $N \to \infty$ one obtains
\begin{eqnarray}\label{third_order_pt}
\lim_{N \to \infty} -\frac{1}{N^2} \ln \tilde F_N\left(x = \frac{L}{\sqrt{2N}} \right) = 
\begin{cases}
&\phi_-^A(x) \;, \; x < 1 \\
&0 \;, \; x > 1 \;.
\end{cases}
\end{eqnarray}
If one interprets $\tilde F_N(L)$ in Eq.~(\ref{Fp}) as the partition function of a discrete 
Coulomb gas, its logarithm can be interpreted as its free energy. Since $\phi_-^A(x) \sim 
(1-x)^3$ when $x$ approaches $1$ from below, then the third derivative of the free energy at 
the critical point $x = 1$ is discontinuous, which can then be interpreted as a third order 
phase transition.

On the other hand, comparing the asymptotic behavior of ${\cal F}_1$ in Eq. 
(\ref{asympt_behaviors}) with the ones of the rate functions (\ref{asympt_rate}) we can check 
that the expansion of the large deviation functions around the transition point coincides 
with the tail behaviors of the central region Tracy-Widom scaling function. This property 
holds here both for the left and the right tails. 
For instance, consider first the left tail in Eq. (\ref{main_results_F}), i.e., when 
$L\ll \sqrt{2N}$. When $L\to \sqrt{2N}$ from below, we can substitute the asymptotic behavior 
of the rate function $\phi_-^A(x)$ from Eq.
(\ref{asympt_rate}) in the first line of Eq. (\ref{main_results_F}). This gives 
\begin{eqnarray}
\tilde F_N(L) \sim \exp{\left(-\frac{2^{5/2}}{3}N^{1/2}(\sqrt{2N}-L)^{3} \right)}\;, \; 1 \ll 
\sqrt{2N} - L\ll \sqrt{2N} \;,
\label{matching_left.1}
\end{eqnarray} 
On the other hand, consider now the second line of Eq. (\ref{main_results_F}) that describes
the central typical fluctuations. When the deviation from the mean is large, i.e.,
$\sqrt{2N}-L \sim O(\sqrt{N})$ (compared to the typical scale $\sim O(N^{-1/6})$), we can 
substitute in the second line of Eq. (\ref{main_results_F}) the left tail asymptotic behavior 
of the Tracy-Widom function ${\mathcal F}_1(t)$ as described in the first line
of Eq. (\ref{asympt_behaviors}) (with $\beta=1$). This gives, 
\begin{eqnarray}
\tilde F_N(L) \sim \exp \left[-\frac{1}{24}\left(2^{11/6} N^{1/6} 
(\sqrt{2N}-L)\right)^3\right]
\label{matching_left.2}
\end{eqnarray}
which, after a trivial rearrangement, is identical to the expression in
Eq. (\ref{matching_left.1}). Thus the left tail of the central region matches
smoothly with the left large deviation function. 
Similarly, on the right side, using the behavior of 
$\phi_+^A(x)$ in Eq. (\ref{asympt_rate}), one finds from Eq. (\ref{main_results_F}), that
\begin{eqnarray}
1 - \tilde F_N(L) \sim \exp{\left(-\frac{2^{15/4}}{3} N^{1/4}(L-\sqrt{2N})^{3/2}\right)} \;, \; 1 \ll L - \sqrt{2N} \ll \sqrt{2N} \;,
\end{eqnarray} 
which matches perfectly with the right tail of the central part described by ${\cal F}_1(t)$ 
(\ref{asympt_behaviors},~\ref{main_results_F}). Hence in this case of model A, the matching 
between the different regimes is similar to the one found in previous studies of large 
deviation formulas associated with the largest eigenvalue of random 
matrices~\cite{NM09,DM06,VMB07,DM08,MV09,BEMN10}.

\vskip 0.3cm

\noindent {\bf Model B:} In the second model we consider periodic boundary 
conditions on the line segment $[0,L]$. Alternatively, one can
think of the domain as a 
circle of circumference $L$ (of radius
$L/2 \pi$). This corresponds to $N$-dimensional Brownian motion in an affine Weyl chamber of type $\tilde A_{n-1}$~\cite{Gra99,Gra02}. All walkers start initially in the vicinity of a point on the 
circle which we call the origin. We can label the positions
of the walkers by their angles $\{\theta_1,\theta_2,\ldots, \theta_N\}$ 
(see section \ref{derivation} for details). 
Let the initial 
angles be denoted by
$\{\epsilon_1,\epsilon_2,\ldots, \epsilon_N\}$  where $\epsilon_i$'s are 
small. Eventually we will take the limit $\epsilon_i\to 0$.  
We denote by 
${R}_L^B(1)$ the reunion probability after time $\tau = 1$ (note
that the walkers, in a bunch, may wind the circle multiple times), i.e,
the probability that the walkers return to their initial positions
after time $\tau=1$ (staying non-intersecting over the time interval $\tau\in 
[0,1]$). Evidently ${R}_L^B(1)$ depends on $N$ and also on the starting
angles $\{\epsilon_1,\epsilon_2,\ldots, \epsilon_N\}$. To avoid
this additional dependence on the $\epsilon_i$'s, let us introduce
the normalized reunion probability defined as the ratio
\begin{equation}
\label{28}
\tilde{G}_N(L)= {{R}_L^B(1) \over {R}_\infty^B(1)} \;,
\end{equation}
where we assume that we have taken the $\epsilon_i\to 0$ limit. 
One can actually check that the $\epsilon_i$'s dependence actually cancels out between the 
numerator and the denominator: this is the main motivation for studying this ratio of reunion 
probabilities (\ref{28}) in this case. Although in the case of model A it is rather natural 
to expect that this limit $\epsilon_i \to 0$ is well defined, given the interpretation of 
$\tilde F_N(L)$ as the cumulative distribution of the maximal height of non-intersecting 
excursions, this is not so obvious for $\tilde G_N(L)$ where such an interpretation does 
not exist. Nevertheless one can show that this property also holds in this case, yielding 
(see Ref. \cite{FMS11} and also section \ref{derivation})

\begin{equation}\label{29}
\tilde{G}_N(L) = \frac{B_N}{L^{N^2}}\sum_{n_1=-\infty}^\infty 
\ldots \sum_{n_N = -\infty}^{\infty} 
\Delta_N^2(n_1,\ldots,n_N) e^{-\frac{2 \pi^2}{L^2} \sum_{j=1}^N n_j^2} \;,
\end{equation} 
where $\Delta_N(n_1,\ldots, n_N)$ is the Vandermonde determinant (\ref{vdm}) and the 
prefactor
\begin{equation}
B_N=\frac{1}{(2\pi)^{N/2-N^2}\prod_{j=0}^{N-1}\Gamma(j+2)}
\label{norm-Ap}
\end{equation}
ensures that $\tilde{G}_N(L\to \infty)=1$. 
In Ref. \cite{FMS11} it was shown that this normalized reunion probability $\tilde{G}_N(L)$ is,
up to a prefactor, exactly identical to the
partition function of the $2$-d Yang-Mills theory on a sphere with 
gauge group ${\rm U}(N)$:
\begin{eqnarray}\label{relation_B_YM}
\tilde G_N(L) \propto {\cal Z}\left( A = \frac{4\pi^2N}{L^2}, {\rm U}(N)\right) \;.
\end{eqnarray} 
This partition function ${\cal Z}\left( A, {\rm U}(N)\right)$ exhibits the 
Douglas-Kazakov third order phase transition for $A = \pi^2$ which means
that in that case the transition between the right tail and the left tail 
behavior of $\tilde G_N(L)$ occurs for $L=L_c(N) = \sqrt{4N}$. 
In this paper, we provide a detailed analysis of the various regimes, 
right tail, central part and left tail of $\tilde G_N(L)$. 
Our results can be summarized as follows
\begin{eqnarray}\label{main_results_G}
\begin{cases}
\tilde G_N(L) &\sim \exp{\left[-N^2 \phi_-^B\left(L/\sqrt{4N} \right) \right]} \;, \; 
L < \sqrt{4N} \; \& \; |L-\sqrt{4N}| \sim {\cal O}(\sqrt{N}) \\
& \\
\tilde G_N(L) &\sim {\cal F}_2\left[2^{2/3} N^{1/6} (L-\sqrt{4N}) \right] \;, \; L \sim 
\sqrt{4N} \; \& \; |L-\sqrt{4N}| \sim {\cal O}(N^{-1/6}) \\
& \\
1 - \tilde G_N(L) & \sim (-1)^N \exp{\left[-N \phi_+^B\left(L/\sqrt{4N} \right) \right]} \;, \; 
L > \sqrt{4N} \; \& \; |L- \sqrt{4N}| \sim {\cal O}(\sqrt{N}) \;,\\
\end{cases}
\end{eqnarray}
where ${\cal F}_2$ is the TW distribution for GUE, whose explicit expression is given 
in (\ref{F2}) and its asymptotic behaviors are given in Eq. (\ref{asympt_behaviors}). The 
rate 
functions $\phi_\pm^B(x) = \frac{1}{2}\phi_\pm^A(x)$ can be computed exactly (see later).
Of
particular interest are their asymptotic behaviors 
when $L \to \sqrt{4 N}$ from below (left tail) and from above (right tail), which are given 
by
\begin{eqnarray}\label{asympt_rate_B}
\phi_-^B(x) &\sim& \frac{8}{3}(1-x)^3 \;, \; x \to 1^- \;, \\
\phi_+^B(x) &\sim& \frac{2^{7/2}}{3}(x-1)^{3/2}\;, \; x \to 1^+ \;. \nonumber 
\end{eqnarray}
Notice also the oscillating sign in the right tail of $\tilde G_N(L)$ in Eq. 
(\ref{main_results_G}) which is not problematic here as $\tilde G_N(L)$ does not have the 
meaning of a cumulative distribution. As explained before for model A in 
Eq.~(\ref{third_order_pt}), the cubic behavior of $\phi_-^B(x)$ when $x$ approaches $1$ from 
below is again the signature of a third order phase transition in this model. On the other 
hand, by comparing the asymptotic behavior of ${\cal F}_2$ in Eq. (\ref{asympt_behaviors}) 
with the one of the rate functions (\ref{asympt_rate_B}) we can check, in this case of model 
B, that the expansion of the large deviation functions around the transition point $L = 
\sqrt{4N}$ coincides with the tail behavior of the TW scaling function, only for the left 
tail.

\begin{figure}
\includegraphics[width = \linewidth]{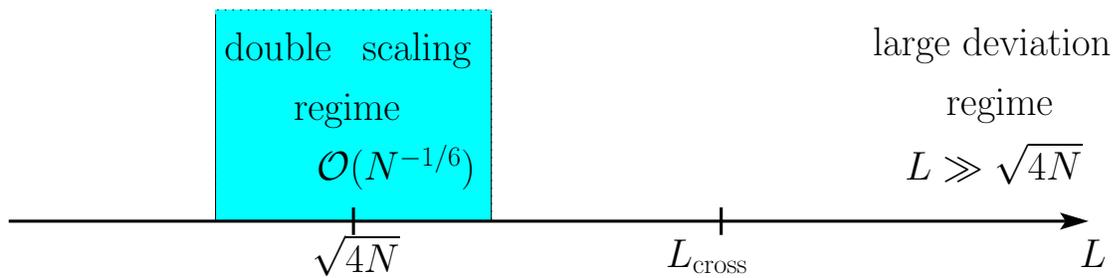}
\caption{Crossover from the double scaling to the large deviation regime which occurs only for model B. This crossover happens for $L_{\rm cross} \sim N^{-1/6} (\ln N)^{2/3}$.}\label{fig_crossover}
\end{figure}

Indeed, using the behavior of $\phi_-^B(x)$ in Eq. (\ref{asympt_rate_B}), one finds from Eq. 
(\ref{main_results_G}) that
\begin{eqnarray}
\tilde G_N(L) \sim \exp{\left[-\frac{1}{3}N^{1/2} (L - \sqrt{4N})^3 \right]} \;, \; 1 \ll 
\sqrt{4N} - L \ll \sqrt{4N}
\end{eqnarray}
which matches perfectly with the left tail of the central region (\ref{asympt_behaviors}). 
However, we find that this property does not hold for the right tail. Indeed, using the 
asymptotic behavior of $\phi_+^B(x)$ in Eq.~(\ref{asympt_rate_B}) one finds from 
(\ref{main_results_G})
\begin{eqnarray}\label{asympt_right_G.0}
1 -\tilde G_N(L) \sim (-1)^N\exp{\left[- \frac{4}{3} N^{1/4}(L - \sqrt{4N})^{3/2}\right]}\;, 
1 \ll L - \sqrt{4N} \ll \sqrt{4N}
\end{eqnarray}
with an oscillating sign $\propto (-1)^N$. On the other hand the right tail of the central 
region, described by ${\cal F}_2(t)$ in Eq. (\ref{asympt_behaviors}), yields for $L - 
\sqrt{4N} \gg 
N^{-1/6}$
\begin{eqnarray}\label{asympt_right_G}
1 - \tilde G_N(L) \sim 1 - {\cal F}_2\left[2^{2/3} N^{1/6} (L-\sqrt{4N}) \right] \sim 
\exp{\left[-\frac{8}{3} N^{1/4} \left(L -  \sqrt{4N} \right)^{3/2} \right]} \;,
\end{eqnarray}
without any oscillating sign and where the argument of the exponential is twice larger than 
the one in (\ref{asympt_right_G.0}). This mismatching is the sign of an interesting 
crossover which happens in this case and which we study in detail below. It can be summarized 
as follows. Close to $L = \sqrt{4N}$, with $L-\sqrt{4N} \sim {\cal O}(N^{-1/6})$, a careful 
computation beyond leading order shows that
\begin{eqnarray}\label{G_crossover}
\ln \tilde G_N(L) = \ln {\cal F}_2(t) + (-1)^{N-1} N^{-1/3} 2^{-1/3} q(t) 
+{\cal O}(N^{-2/3}) \;, \; t = 2^{2/3} N^{1/6} (L-\sqrt{4N}) \;,
\end{eqnarray}  
where $q(t)$ is the Hastings-McLeod solution of PII (\ref{PII}, \ref{asympt_q}). In the large 
$t$ limit, these two competing terms $\ln {\cal F}_2(t)$ and $q(t)$ in Eq. 
(\ref{G_crossover}) behave like (\ref{asympt_q},~\ref{asympt_behaviors})
\begin{eqnarray}\label{asympt_compet}
|\ln {\cal F}_2(t)| \sim e^{-\frac{4}{3}t^{3/2}} \;, \; q(t) \sim {\rm Ai}(t) \sim e^{-\frac{2}{3}t^{3/2}} \;.
\end{eqnarray}
Therefore what happens when one increases $L$ from the critical region $L-\sqrt{4N} \sim 
{\cal O}(N^{-1/6})$ towards the large deviation regime in the right tail $L > \sqrt{4N}$ is 
the following (see Fig. \ref{fig_crossover}): the amplitude of the second term in the right 
hand side of Eq. (\ref{G_crossover}) which is oscillating with $N$, increases relatively to 
the amplitude of the first term. At some crossover value $L \equiv L_{\rm cross}(N)$ it 
becomes larger than the first one and in the large deviation regime it becomes the leading 
term, still oscillating with $N$ (\ref{main_results_G}). Balancing these two terms and making 
use of the above asymptotic behaviors (\ref{asympt_compet}) one obtains an estimate of 
$L_{\rm cross}(N)$ as
\begin{eqnarray}
L_{\rm cross} - \sqrt{4N} \sim N^{-1/6} (\ln N)^{2/3} \;.
\end{eqnarray}  
Note that such a peculiar crossover is absent in the distribution of the largest eigenvalue 
of GUE random matrices and it is thus a specific feature of this vicious walkers problem.

\vskip 0.3cm

\noindent {\bf Model C:} We consider a third model of non-intersecting
Brownian motions where the walkers move again on a finite line segment 
$[0,L]$, 
but this time with {\em reflecting} boundary conditions 
at both boundaries $0$ and $L$. This corresponds to $N$-dimensional Brownian motion in an affine Weyl chamber
of type $\tilde B_N$ \cite{Gra99,Gra02}. Again the walkers start in the 
vicinity of the origin at time 
$\tau=0$ and we consider the reunion probability $R_L^{C}(1)$ that
they reunite at time $\tau=1$ at the origin. Following Models A and B, 
we define the normalized reunion probability 
\begin{equation}\label{RLIII}
\tilde{E}_N(L)={R_L^{C}(1) \over R_\infty^{C}(1)} \;,
\end{equation} 
that is independent of the starting positions $\{\epsilon_1, 
\epsilon_2,\ldots, \epsilon_N\}$ in the limit when all the $\epsilon_i$'s
tend to zero and hence depends only on $N$ and $L$. 
As shown in Ref. \cite{FMS11}, see also in section \ref{derivation}, $\tilde E_N(L)$
can be computed exactly as
\begin{equation}\label{ENL_intro}
\tilde{E}_N(L)= \frac{C_N}{L^{2N^2-N}}\sum_{n_1=-\infty}^\infty
\ldots \sum_{n_N = -\infty}^{\infty}
\Delta^2(n_1^2,\ldots,n_N^2) e^{-\frac{\pi^2}{2L^2} \sum_{j=1}^N n_j^2} \;,
\end{equation}
where $\Delta_N(y_1, \cdots, y_N)$ is the Vandermonde determinant (\ref{vdm.det})
and the prefactor $C_N$ is given by
\begin{equation}
C_N=\frac{{\pi}^{2N^2-N}\,2^{N/2-N^2}}{\prod_{j=0}^{N-1}\Gamma(2+j)\Gamma(1/2+j)} \;,
\label{norm-cn}
\end{equation}
which ensures that $\tilde{E}_N(L\to \infty)=1$. In Ref. \cite{FMS11} we showed that 
$\tilde{E}_N(L)$, up to a prefactor, is exactly identical to the partition function of the 
$2$-d Yang-Mills theory on a sphere with gauge group ${\rm SO}(2N)$:
\begin{eqnarray}\label{relation_C_YM}
\tilde E_{N}(L) \propto {\cal Z}\left(A = \frac{2 \pi^2}{L^2}\, N, {\rm SO}(2N) \right) \;.
\end{eqnarray}
This partition function ${\cal Z}\left( A, {\rm SO}(2N)\right)$ exhibits the Douglas-Kazakov 
third order phase transition for $A = \pi^2$ which means that in that case the transition 
between the right tail and the left tail behavior of $\tilde E_N(L)$ occurs for $L=L_c(N) = 
\sqrt{2N}$. In this paper, we provide a detailed analysis of the various regimes, right tail, 
central part and left tail of $\tilde E_N(L)$. Our results can be summarized as follows
\begin{eqnarray}\label{main_results_E}
\begin{cases}
\tilde E_N(L) &\sim \exp{\left[-N^2 \phi_-^C\left(L/\sqrt{2N} \right) \right]} \;, \; L < \sqrt{2N} \; \& \; |L-\sqrt{2N}| \sim {\cal O}(\sqrt{N})  \\
& \\
\tilde E_N(L) &\sim \dfrac{{\cal F}_2(s)}{{\cal F}_1(s)}\;, \; s = 2^{11/6} N^{1/6} (L-\sqrt{2N})  \;, \; L \sim \sqrt{2N} \; \& \; |L-\sqrt{2N}| \sim {\cal O}(N^{-1/6}) \\
& \\
1 - \tilde E_N(L) & \sim - \exp{\left[-N \phi_+^C\left(L/\sqrt{2N} \right) \right]} \;, \; L > \sqrt{2N} \; \& \; |L-\sqrt{2N}| \sim {\cal O}(\sqrt{N})  \;,\\
\end{cases}
\end{eqnarray}
where ${\cal F}_2$ is the TW distribution for GUE and ${\cal F}_1$ the TW distribution for 
GOE, whose explicit expressions are given in (\ref{F2},~\ref{F2-GOE}) and their asymptotic 
behaviors are given in Eq. (\ref{asympt_behaviors}). The rate functions $\phi_\pm^C(x) = 
\phi_\pm^A(x)$ can be computed exactly (see below): their asymptotic behaviors are given in 
Eq. (\ref{asympt_rate}). As explained before for model A in Eq.~(\ref{third_order_pt}), the 
cubic behavior of $\phi_-^C(x)$ when $x$ approaches $1$ from below is again the signature of 
a third order phase transition in this model. Similarly, as in the case of model A, we can 
check that the expansion of the large deviation functions around the transition point $L = 
\sqrt{2N}$ coincides with the tail behaviors of the central region both in the left tail
\begin{eqnarray}
\tilde E_N(L) \sim - \exp{\left(-\frac{2^{5/2}}{3}N^{1/2}(\sqrt{2N}-L)^{3} \right)}\;, \; 1 
\ll \sqrt{2N} - L \ll \sqrt{2N} \;,
\end{eqnarray}
as well as in the right tail
\begin{eqnarray}
1 - \tilde E_N(L) \sim -\exp{\left(-\frac{2^{15/4}}{3} N^{1/4}(L-\sqrt{2N})^{3/2}\right)} \;, \; 1 \ll L - \sqrt{2N} \ll \sqrt{2N} \;,
\end{eqnarray} 
where, here again, the minus sign is not problematic as $\tilde E_N(L)$ does not have the 
interpretation of a cumulative probability distribution.

\section{Derivation of the formula for the reunion probabilities}\label{derivation}

In this section, we derive the expressions of the normalized reunion probabilities given in 
Eqs.~(\ref{Fp}, \ref{29}, \ref{ENL_intro}). The derivations are based on a 
Fermionic path integral method. For model A, this result was first reported
in Ref.~\cite{SMCR08}.

\subsection{Model A: absorbing boundary conditions at $x=0$ and $x=L$}

We start by the computation of $\tilde F_N(L)$ defined in Eq. (\ref{RL}). 
It then follows that
\begin{eqnarray}\label{ratio_F}
\tilde F_N(L) = \lim_{\epsilon_i \to 0} \left[  \frac{N({\boldsymbol{\epsilon}},L)}{N({\boldsymbol{\epsilon}},L \to \infty)} \right] \;,
\end{eqnarray}
where we use the notation ${\boldsymbol{\epsilon}} \equiv \epsilon_1, \cdots, \epsilon_N$ and 
where $N({\boldsymbol{\epsilon}},L)$ is the probability that the $N$ Brownian paths, with 
diffusion coefficient $D=1/2$, starting at $0 < \epsilon_1< \cdots < \epsilon_N$ at $\tau=0$ 
come back to the same points at $\tau=1$ without crossing each other and staying within the 
interval 
$[0,L]$, with absorbing boundary conditions at $x=0$ and $x=L$. 

To proceed, let us first consider the simple case of $N$ 
{\it independent} and free Brownian walkers on a line, each with diffusion constant $D=1/2$,  
over the unit time interval $[0,1]$, 
{\it but without the  
non-intersection} constraint. The probability measure of an assembly
of $N$ trajectories, that start at $\vec \epsilon$ and also end at $\vec \epsilon$, would 
then be simply proportional to the propagator
\begin{equation}
P_{\rm free}\left(\boldsymbol{\epsilon}, \tau = 0 | \boldsymbol{\epsilon}, \tau=1\right)=
\int {\cal D}x_i(\tau) 
\exp{\left[-\frac{1}{2} \sum_{i=1}^N \int_0^1 \left(\frac{dx_i}{d\tau} \right)^2 
\, d \tau \right]} \prod_{i=1}^N \delta[x_i(0) - \epsilon_i] \delta[x_i(1) - \epsilon_i] \;.
\end{equation} 
If, in addition, we subject the walkers to stay within the box $[0,L]$ during the
time interval $[0,1]$,
this is equivalent to putting an infinite potential at the two ends of the box $[0,L]$.
Then one can use path integral 
techniques to write $N({\boldsymbol{\epsilon}},L)$ as the propagator
\begin{eqnarray}\label{def_HL}
N({\boldsymbol{\epsilon}},L) = \langle {\boldsymbol{\epsilon}} |e^{-\hat H_L}| {\boldsymbol{\epsilon}} \rangle \;, \; H_L = \sum_{i=1}^N \left[-\frac{1}{2} \frac{\partial^2}{\partial x_i^2} + V_L(x_i) \right] \;,
\end{eqnarray}
where $V_L(x)$ is a confining potential with
\begin{eqnarray}
V_L(x) = 
\begin{cases}
&0 \;, \; x \in [0,L] \;, \\
& + \infty \;, \; x \notin [0,L] \;.
\end{cases}
\end{eqnarray}
If one denotes by $E$ the eigenvalues of $\hat H_L$ (\ref{def_HL}) and $| E \rangle$ the 
corresponding eigenvectors, one has
\begin{eqnarray}\label{exp_eigenv}
N({\boldsymbol{\epsilon}},L) = \sum_E |\Psi_E({\boldsymbol{\epsilon}})|^2 e^{-E} \;, \;  
\Psi_E({\boldsymbol{\epsilon}}) = \langle {\boldsymbol{\epsilon}}| E 
\rangle \;.
\end{eqnarray}

So far, we have not implemented the {\em non-intersection} constraint. The
important observation that we make is that this constraint can be incorporated
within the path integral framework by simply insisting that 
the many body wave function 
$\Psi_E({\boldsymbol{x}}) \equiv \Psi_E(x_1, \cdots, x_N)$ must be Fermionic, i.e. it 
vanishes if any of the two coordinates are equal. This many-body antisymmetric wave function 
is thus constructed from the one-body eigenfunctions of $\hat H_L$ by forming the associated 
Slater determinant. In the case of model A, we note that the single particle wave-functions 
vanishing at $x=0$ and $x=L$ are
\begin{eqnarray}\label{phin_A}
\phi_n(x) = \sqrt{\frac{2}{L}} \sin{\left(\frac{n \pi x}{L}\right)} \;, \; n \in \mathbb{Z}^+ \;,
\end{eqnarray}
such that the eigenfunctions $\Psi_E({\boldsymbol{\epsilon}})$ and eigenvalues $E$ in Eq. (\ref{exp_eigenv}) are given here by
\begin{eqnarray}\label{explicit_psi_A}
\Psi_E({\boldsymbol{\epsilon}}) = \frac{1}{\sqrt{N\,!}} \det_{1 \leq i,j \leq N} \phi_{n_i}(\epsilon_j) \;, \; E = \frac{\pi^2}{2L^2} \sum_{i=1} n_i^2 \;.
\end{eqnarray}
Hence one has
\begin{eqnarray}\label{N_explicit}
N(\boldsymbol{\epsilon},L) = \frac{1}{N!}\,\left(\frac{2}{L} \right)^N \sum_{n_1=1}^\infty 
\cdots \sum_{n_N=1}^\infty 
\left|
\begin{array}{cccc}
\sin\left(\frac{n_1 \pi \epsilon_1}{L} \right) & \sin\left(\tfrac{n_1 \pi \epsilon_2}{L} \right) & \cdots & \sin\left(\tfrac{n_1 \pi \epsilon_N}{L} \right) \\
\sin\left(\frac{n_2 \pi \epsilon_1}{L} \right) & \sin\left(\tfrac{n_2 \pi \epsilon_2}{L} \right) & \cdots & \sin\left(\tfrac{n_2 \pi \epsilon_N}{L} \right) \\
\cdot & \cdot & \cdot & \cdot \\
\cdot & \cdot & \cdot & \cdot \\
\sin\left(\frac{n_N \pi \epsilon_1}{L} \right) & \sin\left(\tfrac{n_N \pi \epsilon_2}{L} \right) & \cdots & \sin\left(\tfrac{n_N \pi \epsilon_N}{L} \right) 
\end{array}
\right|^2 e^{-\frac{\pi^2}{2L^2} \sum_{i=1}^N n_i^2}
\end{eqnarray}
We now have to study the limit of $N(\boldsymbol{\epsilon},L)$ in Eq. (\ref{N_explicit}) when 
$\epsilon_i \to 0$ ($\forall i = 1, \cdots, N$). One can then check that to leading order, 
one has
\begin{equation}\label{asympt_det}
\left|
\begin{array}{cccc}
\sin\left(\frac{n_1 \pi \epsilon_1}{L} \right) & \sin\left(\tfrac{n_1 \pi \epsilon_2}{L} \right) & \cdots & \sin\left(\tfrac{n_1 \pi \epsilon_N}{L} \right) \\
\sin\left(\frac{n_2 \pi \epsilon_1}{L} \right) & \sin\left(\tfrac{n_2 \pi \epsilon_2}{L} \right) & \cdots & \sin\left(\tfrac{n_2 \pi \epsilon_N}{L} \right) \\
\cdot & \cdot & \cdot & \cdot \\
\cdot & \cdot & \cdot & \cdot \\
\sin\left(\frac{n_N \pi \epsilon_1}{L} \right) & \sin\left(\tfrac{n_N \pi \epsilon_2}{L} \right) & \cdots & \sin\left(\tfrac{n_N \pi \epsilon_N}{L} \right) 
\end{array}
\right|^2 \sim \frac{a_1}{L^{2N^2}} \left[\prod_{i=1}^N n_i^2 \prod_{i<j} (n_i^2-n_j^2)^2 \right] \left[\prod_{i=1}^N \epsilon_i^2 \prod_{i<j} (\epsilon_i^2-\epsilon_j^2)^2 \right],
\end{equation}
where $a_1$ is a numerical constant independent of $n_i$'s and $\epsilon_i$'s. Therefore one 
obtains
\begin{equation}\label{deriv_f_inter}
\tilde F_N(L) = \frac{a_2}{L^{2N^2 + N}} \lim_{\epsilon_i \to 0} \left[\frac{\prod_{i=1}^N \epsilon_i^2 \prod_{i<j}(\epsilon_i^2 - \epsilon_j^2)^2}{N({\boldsymbol{\epsilon},L \to \infty})} \right] \sum_{n_1=1}^\infty \cdots \sum_{n_N=1}^\infty  \prod_{i=1}^N n_i^2 \prod_{i<j} (n_i^2-n_j^2)^2 e^{-\frac{\pi^2}{2L^2} \sum_{i=1}^N n_i^2} \;,
\end{equation}
with $a_2 = 2^N a_1$. To compute $N({\boldsymbol{\epsilon},L \to \infty})$, starting from Eq. (\ref{N_explicit}) one first notices that the product of determinants can be replaced by the same limiting behavior as in the $\epsilon_i \to 0$ limit given in Eq. (\ref{asympt_det}) as this product is a function of the variables $\epsilon_i/L$'s only. One can then perform the remaining multiple sums over $n_i$'s in (\ref{N_explicit}) by noticing that they can be replaced by integrals in the limit $L \to \infty$ -- as done below in Eq. (\ref{normalisation_1}). Finally, one obtains that 
\begin{eqnarray}\label{limit_epsilon}
\lim_{\epsilon_i \to 0} \left[\frac{\prod_{i=1}^N \epsilon_i^2 \prod_{i<j}(\epsilon_i^2 - \epsilon_j^2)^2}{N({\boldsymbol{\epsilon},L \to \infty})} \right]  = a_3 \;,
\end{eqnarray}
where $a_3$ is a constant independent of $L$. Therefore, combining these results (\ref{deriv_f_inter}, \ref{limit_epsilon}), and using the symmetry of the summand in Eq. (\ref{deriv_f_inter}) under the transformation $n_i \to -n_i$, one arrives at
\begin{eqnarray}\label{deriv_f_inter2}
\tilde F_N(L) = \frac{A_N}{L^{2N^2 + N}} \sum_{n_1=-\infty}^\infty \cdots \sum_{n_N=-\infty}^\infty  \prod_{i=1}^N n_i^2 \prod_{i<j} (n_i^2-n_j^2)^2 e^{-\frac{\pi^2}{2L^2} \sum_{i=1}^N n_i^2} \;,
\end{eqnarray}
as announced before (\ref{Fp}), where the constant $A_N$ remains however undetermined. It can 
be computed using the normalization condition of $\tilde F_N(L)$ which follows directly from 
its definition~(\ref{RL})
\begin{eqnarray}
\lim_{L \to \infty} \tilde F_N(L) = 1 \;.
\end{eqnarray} 
Indeed, as we mentioned it above, in the limit when $L \to \infty$, the discrete variables 
$\pi n_i/L \equiv k_i$ which naturally enter into the expression of $\tilde F_N(L)$ in Eq. 
(\ref{deriv_f_inter2}) become continuous variables. Therefore the 
discrete sums in Eq. (\ref{deriv_f_inter2}) can be replaced by integrals in the $L\to 
\infty$ limit, giving 
\begin{eqnarray}\label{normalisation_1}
\lim_{L \to \infty} \tilde F_N(L) = \frac{A_N}{\pi^{2N^2+N}} \int_{-\infty}^\infty dk_1 \cdots \int_{-\infty}^\infty dk_N \prod_{i=1}^N k_i^2 \prod_{i<j}({k_i^2 - k_j^2})^2 e^{-\frac{1}{2} \sum_{i=1}^N k_i^2} \;.
\end{eqnarray}
If one restricts the integrals over $k_i \in [0,+\infty)$ and performs the change of variable 
$x_i = k_i^2/2$ one obtains
\begin{eqnarray}
\lim_{L \to \infty} \tilde F_N(L) = \frac{A_N}{\pi^{2N^2+N}} 2^{N^2+N/2} \int_0^\infty dx_1 
\cdots \int_0^\infty 
dx_N \prod_{i=1}^N x_i^{1/2} \prod_{i<j} (x_i-x_j)^2 e^{- \sum_{i=1}^N x_i} \;,
\end{eqnarray}
where the integral can now be evaluated using a limiting case of the Selberg integral~\cite{Fo10}:
\begin{eqnarray}\label{selberg}
\int_0^\infty dx_1 \cdots \int_0^\infty dx_N \prod_{i<j} \left(x_i-x_j \right)^{2 \gamma} \prod_{i=1}^N x_i^{\alpha-1} e^{-x_i} = \prod_{j=0}^{N-1} \frac{\Gamma(1 + \gamma+j \gamma) \Gamma(\alpha + \gamma j)}{\Gamma(1+\gamma)} \;,
\end{eqnarray}
yielding (with $\gamma=1$ and $\alpha=3/2$), the formula for $A_N$ given in Eq. 
(\ref{BN_intro}), [see also Eq. (\ref{BN})].

\subsection{Model B: periodic boundary conditions}

Here we consider $N$ non-intersecting Brownian motions with diffusion coefficient $D=1/2$ on 
a line segment $[0,L]$ with periodic boundary conditions or equivalently on a circle of 
radius $L/2\pi$, starting in the vicinity of the origin at time $\tau = 0$. In this case, the 
computation of the ratio of reunion probability in Eq. (\ref{29}) can be done along the same 
line as before, (\ref{def_HL}-\ref{explicit_psi_A}). In this case, the positions of the 
particles are more naturally labelled by their angles $\theta_i$'s (rather than their 
positions $x_i$'s), in terms of which the Hamiltonian $\hat H_L$ (\ref{def_HL}) associated to 
this diffusion reads~(see~\cite{FMS11})
\begin{eqnarray}\label{def_HL_B}
\hat H_L = -\frac{2 \pi^2}{L^2} \sum_{i=1}^N \frac{\partial^2}{\partial \theta_i^2} \;,
\end{eqnarray}
which actually comes from the expression of the bi-dimensional Laplacian in terms of polar variables (we recall that the radius of the circle is $L/2\pi$ and the diffusion coefficient is $D=1/2$). It is easy to see that the one particle eigenfunctions of (\ref{def_HL_B}) $\phi_n(\theta)$ which satisfy the boundary conditions $\phi_n(\theta) = \phi_n(\theta+2 \pi)$ are  given by
\begin{eqnarray}
\phi_n(\theta) = \frac{1}{\sqrt{2 \pi}} e^{i n \theta} \;, n \in \mathbb{N} \;,
\end{eqnarray}
such that the $N$ particle anti-symmetric eigenfunctions $\Psi_E(\boldsymbol{\theta})$ and 
associated eigenvalues $E$ are given by [see Eq. (\ref{explicit_psi_A})]
\begin{eqnarray}
\Psi_E(\boldsymbol{\theta}) = \frac{1}{\sqrt{N!}} \det_{1\leq j,k \leq N} \left[\phi_{n_j}(\theta_k)\right] \;, \; E = \frac{2 \pi^2}{L^2} \sum_{i=1}^N {n_i}^2 \;.
\end{eqnarray}
After some manipulations similar to the one performed before in the case of model A, one 
arrives straightforwardly at the formula for $\tilde G_N(L)$ given in Eq. (\ref{29}).

\subsection{Model C: reflecting boundary conditions at $x=0$ and $x=L$}

In this case we consider $N$ non-intersecting Brownian motions with diffusion coefficient 
$D=1/2$ on the segment $[0,L]$ with reflecting boundary conditions both at $x=0$ and $x=L$. 
The analysis is then exactly the same as the one performed for model A in Eqs. 
(\ref{def_HL})-(\ref{explicit_psi_A}) except that in that case the single particle 
eigenfunctions $\phi_n(x)$ must satisfy: $\partial_x \phi_n(x) = 0$ at $x=0$ and $x=L$. It is 
then easy to see that they are given by
\begin{eqnarray} 
\phi_n(x) = \sqrt{\frac{2}{L}} \cos{\left(\frac{n \pi x}{L}\right)} \;.
\end{eqnarray}
After some manipulations, exactly similar to the one performed in the case of model A, one 
obtains the formula for $\tilde E_N(L)$ given in Eq. (\ref{ENL_intro}).

In the following sections \ref{largeN_A}, \ref{largeN_B}, \ref{largeN_C}, we analyze these 
formula (\ref{Fp}, \ref{29}, \ref{ENL_intro}) in the limit of large $N$.

\section{Large $N$ analysis of the maximal height of $N$ non-intersecting 
excursions}\label{largeN_A}

In this section, we focus on the distribution of the maximal height of $N$ non-intersecting excursions $\tilde 
F_N(L)$ given by
\begin{eqnarray}\label{start.expr.1}
\tilde F_N(L) = \frac{A_N}{L^{2N^2+N}} \sum_{n_1=-\infty}^\infty \cdots  \sum_{n_N=-\infty}^\infty \prod_{i=1}^N n_i^2 \, \Delta_N^2(n_1^2, \cdots, n_N^2) e^{-\frac{\pi^2}{2L^2} \sum_{i=1}^N n_i^2} \;, 
\end{eqnarray}
where $\Delta_N(y_1,\cdots,y_N)$ is the Vandermonde determinant
\begin{eqnarray}\label{vdm.det}
\Delta_N(y_1,\cdots,y_N) = \prod_{1\leq j<k \leq N} (y_j - y_k) \;,
\end{eqnarray}
and the amplitude $A_N$ is given by
\begin{eqnarray}\label{BN}
A_N = \frac{\pi^{2N^2+N}}{2^{N^2 + \frac{N}{2}} \prod_{j=0}^{N-1} \Gamma(2+j) 
\Gamma(\frac{3}{2}+j)} \;.
\end{eqnarray}
We introduce the parameter $\alpha$ defined as
\begin{eqnarray}\label{alpha_absorbing}
\alpha = \frac{\pi^2}{2L^2} \;,
\end{eqnarray}
such that $\tilde F_N(L)$ can be written as
\begin{eqnarray}
&&\tilde F_N(L) = \frac{1}{\prod_{j=0}^{N-1} \Gamma(2+j) \Gamma(\frac{3}{2} + j)} \alpha^{N^2 + \frac{N}{2}} \Omega(\alpha,N) \;, \label{start.expr.2.1}\\
&&\Omega(\alpha,N) = \sum_{n_1=-\infty}^\infty \cdots  
\sum_{n_N=-\infty}^\infty \prod_{i=1}^N n_i^2 \, \Delta^2_N(n_1^2, \cdots, n_N^2) e^{-\alpha \sum_{i=1}^N n_i^2} \;. \label{start.expr.2.2}
\end{eqnarray}

The goal of this section is to provide the large $N$ asymptotic analysis of this formula 
(\ref{start.expr.2.1}, \ref{start.expr.2.2}). In the large $N$ limit, the leading 
contribution to $\tilde F_N(L)$ comes from $L \sim {\cal O}(\sqrt{N})$, and hence $\alpha = 
{\pi^2}/{(2L^2)} \ll 1$. It is thus natural to approximate the multiple sum in 
$\Omega(\alpha,N)$ (\ref{start.expr.2.2}) by a multiple integral and then evaluate this 
multiple integral, in the large $N$ limit, using a saddle point method. This saddle point  
method yields a non trivial result for $L < \sqrt{2N}$, i.e. for the left tail of the 
distribution, and it will be analyzed in subsection \ref{subsect:coulomb}. This calculation 
was done for $YM_2$ with the gauge group $G = {\rm U}(N)$ by Douglas and Kazakov \cite{DK93} 
and we show below that, after simple manipulations, this saddle point analysis in the present 
case where $G = {\rm Sp}(2N)$ can be achieved without any additional calculation, by 
borrowing the results of Ref. \cite{DK93}. On the other hand, for $L>\sqrt{2N}$, the saddle 
point result for $\Omega(\alpha,N)$ in Eq. (\ref{start.expr.2.2}) compensates exactly, to 
leading order, the large $N$ behavior of the prefactor $\alpha^{N^2 + 
\frac{N}{2}}/\prod_{j=0}^{N-1} \Gamma(2+j) \Gamma(\frac{3}{2} + j)$ in Eq. 
(\ref{start.expr.2.1}). Hence for $L > \sqrt{2N}$ the net result of this saddle point 
analysis is simply $\tilde F_N(L) = 1$. The analysis beyond this leading (and trivial) order 
requires the calculation of non-perturbative, instanton-like, contributions to the partition 
function. This was first done by Gross and Matytsin~\cite{GM94} in the case of the gauge 
group $G = {\rm U}(N)$, and then by Crescimanno, Naculich and Schnitzer~\cite{CNS96} for 
other gauge groups including $G = {\rm Sp}(2N)$, using the method of discrete orthogonal 
polynomials. We describe this powerful method in subsection \ref{subsect:ortho_polynomials}, 
which then allows us to analyze the right tail of $\tilde F_N(L)$ in subsection 
\ref{subsect:right}. Finally, this method of orthogonal polynomials also allows us to analyze 
the central part of $\tilde F_N(L)$, which describes the typical fluctuations of the maximal 
height, by studying this system of orthogonal polynomials in the double scaling limit, which 
is done in subsection \ref{subsect.double.scaling}.

\subsection{Coulomb gas analysis for large $N$ and the left tail}\label{subsect:coulomb}

Physically, it is natural to consider the expression of $\Omega(\alpha,N)$ in Eq. 
(\ref{start.expr.2.2}) as the partition function of a discrete Coulomb gas. The large $N$ 
analysis can then be performed along the line of the work of Douglas and Kazakov 
in the context of the partition function of the $YM_2$ with the ${\rm U}(N)$ gauge 
group~\cite{DK93}, 
using a constrained saddle point analysis. As we show below, although the discrete partition sum $\Omega(\alpha,N)$ in our model
is not exactly the same as that of Douglas and Kazakov, it is nevertheless possible to 
transform directly their result to our case without any further additional
work. This then allows us to obtain the left 
tail of the distribution $\tilde F_N(L)$ in the 
large $N$ limit.

For this purpose we first rewrite $\tilde F_N(L)$ in a slightly different way
\begin{eqnarray}
&&\tilde F_N(L) = \frac{2^N}{\prod_{j=0}^{N-1} \Gamma(2+j) \Gamma(\frac{3}{2} + j)} 
\alpha^{N^2 + \frac{N}{2}} \tilde \Omega(\alpha,N) \;, \label{very_start.expr.3} \\
&&\tilde \Omega(\alpha,N) = \sum_{n_1=0}^\infty \cdots  
\sum_{n_N=0}^\infty \prod_{i=1}^N n_i^2 \, \Delta_N^2(n_1^2, \cdots, n_N^2)\, e^{-\alpha 
\sum_{i=1}^N n_i^2} \;, \label{very_start.expr.3.2}
\end{eqnarray} 
where the integers $n_i$ are then all positive $n_i>0$ and recall that $\alpha=\pi^2/{2L^2}$. 
We now regard the summand in Eq. (\ref{very_start.expr.3.2}) as a function of the variables 
\begin{eqnarray}\label{def_x_i}
x_i = \frac{n_i}{2N} \;  {\rm with} \; i=1, \cdots, N \;. 
\end{eqnarray}
In the large $N$ limit the variables $\{n_i/(2N)\}_{i=1,\cdots,N}$ approximate the coordinates 
of a continuous $N$ particles system and associated with the particles is a density $\tilde 
\rho(x)$. Because of the Vandermonde determinant these particles experience hard core 
repulsion and behave as fermions. The configuration with highest density corresponds to the 
case where $n_1 = 1, n_2=2, \cdots, n_N = N$ for which the local density is uniform and given 
by $\tilde \rho_{\rm max} = 2$, the lattice spacing being $1/(2N)$. Therefore the density 
$\tilde \rho(x)$ must satisfy the constraint
\begin{eqnarray}\label{def_rho}
\tilde \rho(x) = \frac{1}{N} \sum_{i=1}^N \langle \delta(x-x_i) \rangle \leq 2 \;, \; \int_0^a dx \, \tilde \rho(x) = 1 \;,
\end{eqnarray} 
where the average $\langle \cdots \rangle$ is taken with respect to the discrete weight in Eqs. 
(\ref{very_start.expr.3},~\ref{very_start.expr.3.2}) and where $[0,a]$ is thus the support of 
the mean density, $a$ remaining to be determined. 

The main idea is that in the large $N$ limit the discrete sum can be replaced by a multiple
integral over continuous variables and this multiple integral can be viewed
as the partition function of a Coulomb gas with $x_i=n_i/{2N}$ denoting
the position of the $i$-th charge. The next step is to replace this
multiple integral over $N$ variables by a functional integral
over the coarse grained density $\tilde \rho(x)$ \`a la Dyson \cite{Dy62a,Dy62b,Dy62c} (for a review see Refs.~\cite{DM08,Fo10}).
Skipping details, one finds that
in terms of the scaling variable
\begin{eqnarray}\label{def.h}
h = \frac{L}{\sqrt{2N}} \;,
\end{eqnarray}
one can approximate to leading order in the large $N$ limit,
\begin{eqnarray}\label{expr_cg}
&&\tilde F_N(h \sqrt{2N}) \sim 
\int {\cal D} \tilde \rho(x) e^{-N^2 S[\tilde \rho]} \;, \\
&&S[\tilde \rho] = \frac{\pi^2}{h^2} \int_0^a dx x^2 \tilde \rho(x) - 
\int_0^a dx \int_0^a dx' \tilde \rho(x) \tilde \rho(x') \ln {|x^2 - x'^2|}
+ C\, \left[\int_0^{a} \tilde \rho(x)\, dx -1\right]
\label{def_S}
\end{eqnarray}
where $C$ is a Lagrange multiplier that enforces the normalization condition
of the charge density. 
If we introduce $\rho(x) = \tilde \rho(x)/2$, for $x>0$ and $\rho(x) = \tilde \rho(-x)/2$ 
for $x<0$ it is easy to see that 
\begin{eqnarray}
&&S[\tilde \rho] = 2 S_{DK}[\rho] \\
&& S_{DK}[\rho]= \frac{\pi^2}{2h^2} \int_{-a}^a dx x^2 \rho(x) - 
\int_{-a}^a dx \int_{-a}^a dx' \rho(x) \rho(x') \ln{|x - x'|}
+ C' \left[\int_{-a}^a \rho(x) dx -1\right]
 \;,
\end{eqnarray}
where $C'=C/2$ is a constant and 
$S_{DK}[\rho]$ is exactly the action studied by Douglas and Kazakov \cite{DK93} with 
the substitution $h^2 = \pi^2/A$. 

In the large $N$ limit, the path integral over $\tilde 
\rho$ in Eq. (\ref{expr_cg}) can be evaluated by the saddle point method, giving 
\begin{eqnarray}\label{formule_SP}
\int {\cal D} \tilde \rho(x) e^{-N^2 S[\tilde \rho]} \sim \exp{\left(-2N^2 S_{DK}[\rho^*]\right)} \;,
\end{eqnarray} 
where $\rho^*$ is such that
\begin{eqnarray}\label{stationary}
\frac{\delta S_{DK}[\rho]}{\delta \rho(x)} \Bigg |_{\rho=\rho^*} = 0 \;,
\end{eqnarray}
which gives an integral equation for the saddle point density 
\begin{eqnarray}
\frac{\pi^2}{2h^2}\, x^2 - 2 \int_{-a}^a \rho^*(x')\ln |x-x'|\, dx' + C'=0
\label{saddle.1}
\end{eqnarray}
that holds only over the support $x\in [-a,a]$ where $\rho^*(x)$ is nonzero.
Taking a further derivative with respect to $x$ gets rid of the Lagrange multiplier $C'$
and leads to a singular integral equation for $\rho^*(x)$ 
\begin{eqnarray}\label{saddle_point_eq}
\frac{\pi^2}{2h^2} x - \dashint_{-a}^a \frac{\rho^*(x')}{x-x'} dx' = 0 \;, 
\end{eqnarray}
where $\dashint$ stands for the principal part of the integral, together with the constraints 
on $\rho(x)$, inherited from the ones on $\tilde \rho$ in Eq. (\ref{def_rho})
\begin{eqnarray}\label{constraint_rho}
\int_{-a}^a \rho^*(x) dx = 1 \;, \; \rho^*(x) \leq 1 \;, \forall x \in [-a,a] \;.
\end{eqnarray}

If one first discards the constraint $\rho^*(x) \leq 1$, the solution of the saddle point 
equation (\ref{saddle_point_eq}) is simply given by the Wigner semi-circle law \cite{Fo10}
\begin{eqnarray}\label{semi_circle}
\rho^*(x) = \frac{\pi}{2h^2} \sqrt{\frac{4h^2}{\pi^2} - x^2} \:,
\end{eqnarray}
and thus $a = 2h/\pi$. 
\begin{figure}
\includegraphics[width = \linewidth]{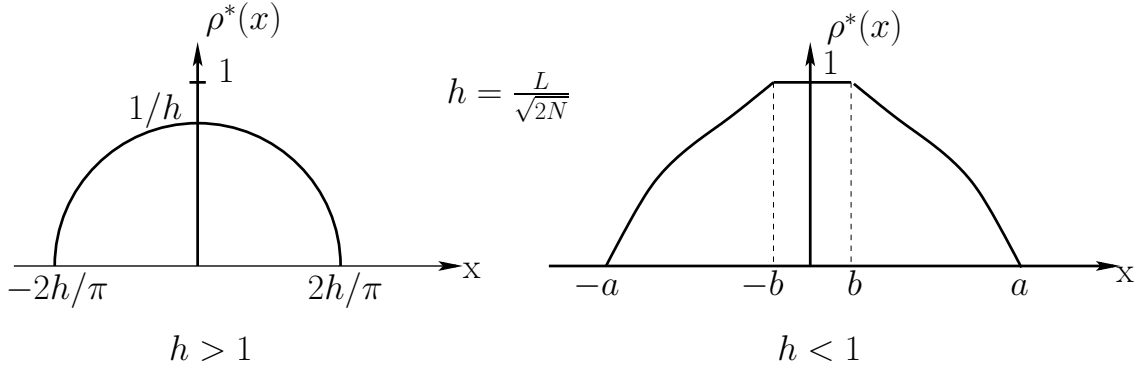}
\caption{Sketch of the density $\rho^*(x)$ for $h<1$ (left tail), and $h>1$ (right tail). The Douglas-Kazakov transition occurs for $h=1$, when $L = L_c(N)=\sqrt{2N}$ in model A, where $\rho^*(0) = 1$.}\label{fig_dk}
\end{figure}
This density $\rho^*(x)$ has its maximum at $x=0$, where $\rho^*(0) = 1/h$: therefore 
the Wigner semi-circle law is the solution of this constrained saddle point equation as long 
as $h>1$. Noting that $h= L/\sqrt{2N}>1$ means $L>\sqrt{2N}$, this
solution then corresponds to the 
right tail of the distribution $\tilde F_N(L)$ (see Fig. 
\ref{fig_dk}).

What happens when $h<1$, or equivalently $L<\sqrt{2N}$? Clearly, the Wigner semi-circle is
no longer the right solution since it violates the constraint $\rho^*(x)\le 1$. So,
there must be another solution to the singular integral equation when $h<1$.
When $h$ approaches $1$ from above, the maximum of the semi-circle at $x=0$ just touches 
$1$.
As one decreases $h$ below $1$, the density at $x=0$ cannot obviously increase beyond 
$1$ since its maximum value is $1$. What happens instead is like a `wetting' transition,
i.e., the density 
becomes flat and equal to $1$ symmetrically  
around $x=0$ over an interval $[-b,b]$ and outside the interval it has a nontrivial
shape (see the right panel of Fig. \ref{fig_dk}). In other words, the solution reads
\begin{eqnarray}\label{explicit_rho1}
\rho^*(x) = 1 \;, x \in [-b,b] \;, \rho^*(x) = \hat \rho(x) \;,\; x \in [-a,-b] \cup [a,b] \;,
\end{eqnarray}
where $\hat \rho(x)$ is non zero only for $x \in [-a,-b] \cup [a,b]$.
Substituting this ansatz in Eq. (\ref{saddle_point_eq})
provides an integral equation for the nontrivial part $\hat \rho(x)$ of the density
\begin{eqnarray}\label{integ_rho_tilde}
\frac{\pi^2}{2h^2} x - \ln{\left(\frac{x-b}{x+b} \right)} - 
\dashint_{-a}^a \frac{\hat \rho(x')}{x-x'} dx' = 0 \;.
\end{eqnarray}
Note that both $a$ and $b$ have to be self-consistently determined.
This integral equation for $\hat \rho$ can be solved explicitly in 
terms of elliptic functions \cite{DK93} (here we give the expression from Ref. \cite{GM95}):
\begin{eqnarray}\label{explicit_rho2}
\hat \rho(x) = \frac{2}{\pi a x}\sqrt{(a^2-x^2)(x^2-b^2)} \Pi_1\left(-\frac{b^2}{x^2}, \frac{b}{a}\right) \;,\; x \in [-a,-b] \cup [a,b] \;, 
\end{eqnarray}
where $\Pi_1(y,k)$ is the elliptic integral of the third kind, which admits the following integral representation
\begin{eqnarray}
\Pi_1(y,k) = \frac{1}{2} \int_{-1}^1 \frac{dz}{1+y\,z^2} \frac{1}{\sqrt{1-k^2z^2}\sqrt{1-z^2}} \;.
\end{eqnarray} 
The constants $a$ and $b$ in Eqs (\ref{explicit_rho1}, \ref{explicit_rho2}) can be expressed in terms of standard elliptic integrals (using the notations of Ref. \cite{erdelyi})
\begin{eqnarray}
{\bf K}(y) = \int_0^1 \frac{dz}{\sqrt{1-y^2 z^2}\sqrt{1-z^2}} \;, \; {\bf E}(y) = \int_0^1 dz \frac{\sqrt{1-y^2 z^2}}{1 - z^2} \;.
\end{eqnarray}
Then $a$ and $b$ are determined by the following relations
\begin{eqnarray}
&&k = \frac{b}{a} \label{relation_saddle.1} \\
&&a [2{\bf E}(k)-(1-k^2){\bf K}(k)] = 1  \label{relation_saddle.2} \\
&&a \frac{\pi^2}{h^2} = 4{\bf K}(k) \label{relation_saddle.3} \;.
\end{eqnarray} 

Reverting back to our original problem, described by the action $S[\tilde \rho]$ in Eq. 
(\ref{expr_cg}), with $\tilde \rho(x) = 2 \rho(x)$, we get that the saddle point is obtained 
for $\tilde \rho(x)= \tilde \rho^*(x)$ where, for $h>1$,

\begin{eqnarray}
\tilde \rho^*(x) = 
\begin{cases}
& 0 \;, \; x < 0 \\
&\dfrac{\pi}{h^2} \sqrt{\dfrac{4h^2}{\pi^2} - x^2} \:, \; 0 \leq x \leq \dfrac{2h}{\pi} \\
& 0 \;, \; x > \dfrac{2h}{\pi}
\end{cases}
\end{eqnarray}
while for $h<1$ one has  
\begin{eqnarray}
\tilde \rho^*(x) =
\begin{cases}
&0  \;, \; x < 0 \\
& 2 \;,\; 0 \leq x \leq b \\
& \dfrac{4}{\pi a x}\sqrt{(a^2-x^2)(x^2-b^2)} \Pi_1\left(-\dfrac{b^2}{x^2}, \dfrac{b}{a}\right) \;, \; b \leq x \leq a \\
& 0 \;, \; x > a \;. 
\end{cases}
\end{eqnarray}

These explicit expressions for $\rho^*(x)$ given in Eq. (\ref{semi_circle}) for $h>1$ and in 
Eqs.  (\ref{explicit_rho1},~\ref{explicit_rho2}) for $h<1$ can then be plugged into Eq. 
(\ref{formule_SP}) as done in Ref. \cite{DK93} to obtain
%
\begin{eqnarray}\label{main_res_CG}
\lim_{N \to \infty} -\frac{1}{N^2} \ln \tilde F_N(h\sqrt{2N}) = 
\begin{cases}
0 \;, \; h \geq 1 \\
\phi_A^-(h) = 2 \left(F_-(\pi^2/h^2) - F_+(\pi^2/h^2)\right) \;, \; h < 1 \;,
\end{cases}
\end{eqnarray} 
where 
\begin{eqnarray}
F_-(X) = - \frac{3}{4} - \frac{X}{24} - \frac{1}{2} \ln X \;,
\end{eqnarray}
while $F_+(X)$ has a more complicated, though explicit, expression in terms of elliptic integrals:
\begin{eqnarray}
F_+'(X) = \frac{a^2}{6} - \frac{a^2}{12} (1-k^2) - \frac{1}{24} + \frac{a^4}{96} (1-k^2)^2 X \;,
\end{eqnarray}
where $a$ and $k$ are defined by the relations in Eqs. (\ref{relation_saddle.1})-(\ref{relation_saddle.3}) with the substitution, in the last relation (\ref{relation_saddle.3}) $\pi^2/h^2 \to X$. This result in Eq. (\ref{main_res_CG}) yields the expression of the left tail of $\tilde F_N(L)$ announced in the section 2 in Eq. (\ref{main_results_F}). 

The result of this Coulomb gas approach in Eq. (\ref{main_res_CG}) indicates that this quantity $\tilde F_N(L)$ exhibits a phase transition at $h=1$, i.e. at $L = L_c(N) = \sqrt{2N}$. To investigate the order of this transition, we need to analyze the behavior of the rate function $\phi^-_A(h)$ in Eq. (\ref{main_res_CG}) when $h\to 1$, which requires the behavior of $F_-(X) - F_+(X)$ for $X$ close to $\pi^2$, which was computed in Ref. \cite{DK93} as \footnote{Note that we have corrected a sign error from Ref. \cite{DK93} and reproduced also in Ref. \cite{FMS11}}
\begin{eqnarray}\label{third_order}
F_-(X) - F_+(X) = \frac{1}{3\pi^6} \left({X-\pi^2}\right)^3 + {\cal O}((X-\pi^2)^4) \;.
\end{eqnarray}
This cubic behavior implies a third order phase transition as the third derivative of the rate function presents a discontinuity when $h$ crosses the critical value $h=1$. Note also that, when $h \to 1$, $h<1$, one obtains that $\tilde F_N(h \sqrt{2N})$ behaves like
\begin{eqnarray}
\tilde F_N(h \sqrt{2N}) \sim e^{- N^2\frac{16}{3}(1-h)^3} \;,
\end{eqnarray}
which yields the asymptotic behavior of the rate function $\phi_A^-(h)$ given in Eq.~(\ref{asympt_rate}).

We see in Eq. (\ref{main_res_CG}) that this saddle point approach does not say 
anything 
relevant about the right tail of the distribution. 
It just gives the trivial leading order result that $\tilde F_N(L)\approx 1$ for
$L>\sqrt{2N}$.
To investigate this regime beyond this trivial leading order, we have to
find some other method. Fortunately such a method exists and it involves
discrete orthogonal polynomials which were introduced and studied by Gross and Matytsin 
\cite{GM94} in the context of $YM_2$ with the ${\rm U}(N)$ gauge group. We
can adapt their method to our problem as shown in the next subsection. 
We will see that this method also allows us to study not just the
extreme right tail but also the typical fluctuations, 
for 
$L  \sim \sqrt{2 N}$, which is described by a double scaling limit which we analyze below. 

\subsection{General expression in terms of discrete orthogonal polynomials}\label{subsect:ortho_polynomials}

To analyze (\ref{start.expr.2.1},~\ref{start.expr.2.2}) in the large $N$ limit, we 
follow Ref.~\cite{GM94} and introduce 
the following discrete orthogonal polynomials
\begin{eqnarray}\label{ortho_condition}
\sum_{n=-\infty}^\infty p_k(n|\alpha) p_{k'}(n|\alpha) e^{-\alpha n^2} = \delta_{k,k'} h_k(\alpha) \;, 
\end{eqnarray}
where we recall that $\alpha = \pi^2/2L^2$. Here $p_k(n|\alpha)$'s are monic polynomials of 
degree $k$:
\begin{eqnarray}\label{def.monic}
p_k(n|\alpha) = n^k + \cdots \;
\end{eqnarray}
and $h_k(\alpha)$'s are to be determined. 

These orthogonal polynomials turn out to be very useful as
they allow us to express $\tilde F_N(L)$ in a 
rather compact form, as we show now. First notice that
\begin{eqnarray}\label{expand_vdm}
\prod_{i=1}^n n_i \, \Delta(n_1^2, n_2^2, \cdots, n_N^2) = 
\det_{1\leq i,j \leq N} \left[p_{2i-1} (n_j)\right] \;.
\end{eqnarray}
Using this in Eq. (\ref{start.expr.2.2}) we get
\begin{eqnarray}
\Omega(\alpha,N) = 
\sum_{n_1=-\infty}^\infty \cdots  \sum_{n_N=-\infty}^\infty 
e^{-\alpha \sum_{i=1}^N n_i^2} 
\left[\det_{1\leq i,j \leq N} \left[p_{2i-1} (n_j)\right]\right]^2 \,. 
\label{omega.orthogonal} 
\end{eqnarray}
Next, using the definition in Eq. (\ref{ortho_condition}) and 
the discrete version of the Cauchy-Binet formula one gets
\begin{eqnarray}
\Omega(\alpha,N) = \Gamma(N+1) \prod_{j=1}^{N} h_{2j-1}(\alpha) \;.
\label{omega.orthogonal.2}  
\end{eqnarray}
Thus, if we can determine the $h_k(\alpha)$'s, in principle we can determine $\Omega(\alpha,N)$ exactly.

So, our next task is to determine the $h_k(\alpha)$'s. As we show below, one can actually
set up a very nice recursive procedure to determine the $h_k(\alpha)$'s.
To proceed, we notice that the two first polynomials are obtained straightforwardly~as
\begin{eqnarray}\label{p01}
p_0(n|\alpha) = 1 \;, \; p_1(n|\alpha) = n \;,
\end{eqnarray}
from which one gets the first amplitudes
\begin{eqnarray}\label{h01}
h_0(\alpha) = \sum_{n = -\infty}^\infty e^{-\alpha n^2} \;, \; h_1(\alpha) = \sum_{n=-\infty}^\infty n^2 e^{-\alpha n^2} \;.
\end{eqnarray}
Note that given that the weight $e^{-\alpha n^2}$ is symmetric under the transformation $n \to -n$ and given that the sum in Eq. (\ref{ortho_condition}) runs over $n \in \mathbb{Z}$, one can show (for instance by induction over $k$)  that
\begin{eqnarray}\label{parity}
&&p_{2k}(n) = n^{2k} + a_{2k-2} n^{2k-2} + \cdots + a_0 \\
&&p_{2k+1}(n) = n^{2k+1} + b_{2k-1} n^{2k-1} + \cdots + b_1 n \;. \nonumber
\end{eqnarray}
Because $p_k$'s are orthogonal polynomials (\ref{ortho_condition}) they satisfy a three term recursion relation~\cite{szego}:
\begin{eqnarray}\label{recursion.gen}
n p_k(n |\alpha) = p_{k+1}(n|\alpha) + S_k(\alpha) p_k(n|\alpha) + R_k(\alpha) p_{k-1}(n|\alpha) \;.
\end{eqnarray}
Multiplying the above recursion relation (\ref{recursion.gen}) by $p_k(n|\alpha)e^{-\alpha n^2}$ and summing over $n$, one obtains from (\ref{parity}) that 
\begin{equation}
S_k(\alpha) = 0 \;. 
\end{equation}
On the other hand, multiplying both sides of Eq. (\ref{recursion.gen}) by $p_{k-1}(n|\alpha)e^{-\alpha n^2}$ and summing over $n$ one obtains
\begin{eqnarray}\label{Rk}
R_k(\alpha) = \frac{h_k(\alpha)}{h_{k-1}(\alpha)} \;.
\end{eqnarray}
From Eq. (\ref{h01}), one has 
\begin{eqnarray}\label{R1}
R_1(\alpha) = \frac{h_1(\alpha)}{h_0(\alpha)}= \frac{\sum_{n=-\infty}^\infty n^2 e^{-\alpha n^2}}{\sum_{n=-\infty}^\infty e^{-\alpha n^2}} = -\frac{d}{d\alpha} \ln h_0(\alpha) \;.
\end{eqnarray}
Inspired by the relation found above for $R_1(\alpha)$ (\ref{R1}), we differentiate the orthogonality condition~(\ref{ortho_condition}) with respect to $\alpha$
\begin{eqnarray}\label{differentiate.1}
\frac{d}{d\alpha} h_k(\alpha) = - \langle n p_k|n p_k\rangle + 2 \langle p_k|\partial_\alpha p_k\rangle \;,
\end{eqnarray}
where we have introduced the notation $\langle f | g\rangle$, for two functions $f, g$:
\begin{eqnarray}
\langle f | g\rangle = \sum_{n=-\infty}^\infty f(n) g(n) e^{-\alpha n^2} \;.
\end{eqnarray}
Using the fact that $p_k$ is a monic polynomial (\ref{def.monic}), $\partial_\alpha p_k$ is thus a polynomial of degree at most $k-2$ (\ref{parity}). And therefore, from the orthogonality condition (\ref{ortho_condition}), the second term in (\ref{differentiate.1}) vanishes. Finally, using the recursion relation (\ref{recursion.gen}) one obtains
\begin{eqnarray}\label{differentiate.2}
\frac{d}{d\alpha} h_k(\alpha) = - \left(R_{k+1}(\alpha) h_k(\alpha) + R_k(\alpha) h_k(\alpha) \right) \;.
\end{eqnarray}
From the relation found above for $R_1(\alpha)$ (\ref{R1}) we see, from Eq. (\ref{differentiate.2}) that one has, for consistency:
\begin{equation}\label{R0}
R_0(\alpha) = 0 \;.
\end{equation}
By subtracting the relations (\ref{differentiate.2}) for $\frac{d}{d\alpha} h_k(\alpha)$ and $\frac{d}{d\alpha} h_{k-1}(\alpha)$ one obtains
\begin{eqnarray}\label{volterra}
- \frac{d}{d\alpha} \ln R_k(\alpha) = R_{k+1}(\alpha) - R_{k-1}(\alpha) \;.
\end{eqnarray}

Let us then summarize the main program to be carried out.
We have a set of variables $R_k(\alpha)$'s that satisfy
a Volterra type recursion relation in Eq. (\ref{volterra}), starting
from the initial values: $R_0(\alpha)=0$ and $R_1(\alpha)= -\frac{d}{d\alpha} \ln 
h_0(\alpha)$ where $h_0(\alpha)= \sum_{n=-\infty}^{\infty} e^{-\alpha n^2}$.
This is a closed set of deterministic recursive relations, albeit nonlinear.
In principle, if we have the solution $R_k(\alpha)$ for arbitrary $k$, then we can determine 
the $h_k(\alpha)$'s by iterating the relation (\ref{Rk}): $h_k(\alpha)= R_k(\alpha) 
h_{k-1}(\alpha)$, starting from $h_0(\alpha)= \sum_{n=-\infty}^{\infty} e^{-\alpha n^2}$.
Once we have the $h_k(\alpha)$'s for all $k$, by taking their product in Eq. 
(\ref{omega.orthogonal.2}),
we can determine, at least in principle, $\Omega(\alpha,N)$  and hence subsequently $\tilde 
F_N(L)$ using Eq. (\ref{start.expr.2.1}).

Actually, to get rid of the complicated prefactor in Eq. (\ref{start.expr.2.1}), it turns out 
to be useful to consider the following ratio  
\begin{eqnarray}\label{start.expr.3}
\frac{\tilde F_{N+1}(L) \tilde F_{N-1}(L)}{\tilde F^2_N(L)} = 
\frac{\alpha^2}{N(N+1/2)} \frac{h_{2N+1}(\alpha)}{h_{2N-1}(\alpha)}\;  
=  \frac{\alpha^2}{N(N+1/2)}  R_{2N+1}(\alpha) R_{2N}(\alpha) \;,
\end{eqnarray}
where we have used Eqs. (\ref{start.expr.2.1}), (\ref{omega.orthogonal.2})
and (\ref{Rk}) to simplify. This will turn out to be the most useful form
on which one can perform the asymptotic analysis for large $N$ and large $L$.

So, essentially, our aim and the main challenge are to solve the nonlinear recursion relation 
for the $R_k(\alpha)$'s
in Eq. (\ref{volterra}). Evidently, this can not be solved for all $k$ exactly.
However, for large $k$ and large $L$ (i.e., small $\alpha=\pi^2/{2L^2}$), it turns
out that one can make some scaling ansatz and solve the resulting differential
equation for the scaling functions. This provides us a handle on
the asymptotic analysis of $\tilde F_N(L)$, as we show below in the next two subsections.

\subsection{Large deviation regime: right tail}\label{subsect:right}

To begin with, we first consider the limit when $L \to \infty$, {\it i.e.} $\alpha = \frac{\pi^2}{2L^2}\to 0$, with $N$ fixed to a large value. To get insight into this limit, it is useful to study the amplitudes $h_0(\alpha), h_1(\alpha)$ in Eq. (\ref{h01}) when $\alpha \to 0$. For this purpose, we use the Poisson summation formula: if for a given function $f(x)$ we define its Fourier transform
\begin{eqnarray}\label{poisson.1}
\hat f(q) = \frac{1}{2\pi}\int_{-\infty}^\infty f(x) e^{i q x} \, dx \;,
\end{eqnarray}
then one has
\begin{eqnarray}\label{poisson.2}
\sum_{n=-\infty}^\infty \hat f(n) = \sum_{n=-\infty}^\infty f(2\pi n) \;.
\end{eqnarray}
When applied to $h_0(\alpha)$ and $h_1(\alpha)$ this Poisson formula yields
\begin{eqnarray}
&&h_0(\alpha) = \sqrt{\frac{\pi}{\alpha}} \sum_{n=-\infty}^\infty e^{-\frac{\pi^2 n^2}{\alpha}} = \sqrt{\frac{\pi}{\alpha}} \left(1 + 2 \, e^{-\frac{\pi^2}{\alpha}}  + {\cal O}(e^{-\frac{4\pi^2}{\alpha}})\right) \;, \\
&&h_1(\alpha) = \frac{\sqrt{\pi}}{2 \alpha^{\frac{5}{2}}} \sum_{n=-\infty}^\infty e^{-\frac{\pi^2 n^2}{\alpha}} \left(-2\pi^2 n^2 +  \alpha \right) = \frac{\sqrt{\pi}}{2 \alpha^{\frac{3}{2}}} \left(1  - \left(4 \frac{\pi^2}{\alpha}-2\right) e^{-\frac{\pi^2}{\alpha}}  + {\cal O}\left(\frac{e^{-\frac{4\pi^2}{\alpha}}}{\alpha}\right) \right) \;, \nonumber 
\end{eqnarray}
from which one gets
\begin{eqnarray}\label{r1.asympt}
R_1(\alpha) = \frac{1}{2 \alpha} - \frac{2 \pi^2}{\alpha^2} e^{-\frac{\pi^2}{\alpha}} +  {\cal O}(e^{-\frac{4\pi^2}{\alpha}}/\alpha) \;.
\end{eqnarray}
More generally, one can show that, to leading order, $R_k(\alpha)$ is given by
\begin{eqnarray}\label{rk.leading}
R_k(\alpha) \simeq \frac{k}{2 \alpha} \;, 
\end{eqnarray}
which indeed satisfies the above Eq. (\ref{volterra}). Motivated by the result for $R_1(\alpha)$ (\ref{r1.asympt}) we assume that, when $\alpha \to 0$, the leading correction to (\ref{rk.leading}) is of the form
\begin{eqnarray}\label{rk.ansatz.largedev}
R_k(\alpha) = \frac{k}{2 \alpha} + c_k(\alpha) e^{-\frac{\pi^2}{\alpha}} + {\cal O}(e^{-\frac{4\pi^2}{\alpha}}/\alpha) \;,
\end{eqnarray}
such that $c_k(\alpha) e^{-\frac{\pi^2}{\alpha}} \ll 1$. Note that from (\ref{R0}) and (\ref{r1.asympt}) one has
\begin{eqnarray}\label{c01}
c_0(\alpha) = 0 \;, \; c_1(\alpha) = - 2 \frac{\pi^2}{\alpha^2} \;.
\end{eqnarray}
By inserting this ansatz (\ref{rk.ansatz.largedev}) into the above recursion relation satisfied by $R_k(\alpha)$ (\ref{volterra}), assuming that $c_k(\alpha) e^{-\frac{\pi^2}{\alpha}} \ll 1$, one finds the following recursion relation for $c_k(\alpha)$:
\begin{eqnarray}\label{recursion.ck}
c_{k+1}(\alpha) - c_{k-1}(\alpha) = - \frac{2}{k} c_k(\alpha) - \frac{2\alpha}{k} c'_k(\alpha) - \frac{2\alpha}{k} \frac{\pi^2}{\alpha^2} c_k(\alpha) \;.
\end{eqnarray}
Following Ref. \cite{GM94} we introduce the new variable 
\begin{eqnarray}\label{def_xi}
\xi = \frac{\pi^2}{\alpha} 
\end{eqnarray}
and rewrite
\begin{eqnarray}\label{eq:ck_gk}
c_k(\alpha) = - \frac{2}{\pi^2} \xi^2 G_k(\xi) \;,
\end{eqnarray}
such that $G_k(\xi)$ satisfies the recurrence relation, obtained from (\ref{recursion.ck}),
\begin{eqnarray}\label{recurrence.gk}
G_{k+1}(\xi) - G_{k-1}(\xi) = \frac{2}{k} \left[\xi G_k'(\xi) + G_k(\xi) - \xi G_k(\xi) \right] \;.
\end{eqnarray}
From Eqs. (\ref{c01}) and (\ref{recurrence.gk}) one obtains explicit expressions for the first $G_k$'s as 
\begin{eqnarray}\label{first.gk}
G_0(\xi) = 0 \;, \; G_1(\xi) = 1 \;, G_2(\xi) = -2 \xi +2 \;, G_3(\xi) = 2 \xi^2 - 6 \xi +3 \;.
\end{eqnarray}
This recursion relation (\ref{recurrence.gk}) can be solved by introducing the generating function 
\begin{eqnarray}\label{def.gf.g}
\tilde G(\xi,z) = \sum_{k=1}^\infty G_k(\xi) \, z^k  \;,
\end{eqnarray}
which satisfies the partial differential equation (pde)
\begin{eqnarray}\label{pde.tildeg}
(1-z^2) \frac{\partial \tilde G}{\partial z} - 2 \xi \frac{\partial \tilde G}{\partial \xi} = \left(z+\frac{1}{z} - 2(\xi-1) \right)\tilde G \;,
\end{eqnarray}
where we have used the short notation $\tilde G \equiv \tilde G(\xi,z)$. This pde (\ref{pde.tildeg}) admits the following exact solution~\cite{GM94} 
\begin{eqnarray}\label{exact.gf.g}
\tilde G(\xi,z) = \frac{z}{(1-z)^2} \exp{\left(-\frac{2 \xi z}{1-z} \right)} \;,
\end{eqnarray}
from which $G_k(\xi)$ is readily obtained as
\begin{eqnarray}\label{contour.gk}
G_k(\xi) = \oint_{C_z} \frac{dz}{2\pi i} \frac{1}{z^{k+1}} \tilde G(\xi,z) = \oint_{C_t} \frac{dt}{2\pi i} e^{-2 \xi t} \left( 1 + \frac{1}{t} \right)^k \;,
\end{eqnarray}
where we have performed the change of variable $t = z/(1-z)$.  In the above expression the contour $C_z$ encircles the origin $z=0$ in such a way that $z=1$ is outside of the contour while the contour $C_t$ encircles the origin $t=0$ and passes to the right of $t=-1$. One recognizes from Eq.~(\ref{contour.gk}), for instance by expanding $(1+1/t)^k$ in powers of $1/t$ and integrating term by term that $G_k(\xi)$ can be written as
\begin{eqnarray}\label{gk.laguerre}
G_k(\xi) = \sum_{m=0}^{k-1} \frac{(-1)^m}{m!} \frac{k!}{(m+1)!(k-m-1)!} (2 \xi)^k = L_{k-1}^{(1)}(2 \xi) \;,
\end{eqnarray}
where $L_{k-1}^{(1)}(x)$ are generalized Laguerre polynomials~\cite{szego}. Therefore one gets, coming back to the variable $\alpha = \frac{\pi^2}{2L^2}$ (\ref{def_xi}), when $\alpha \to 0$ 
\begin{eqnarray}\label{rk.largedev.explicit}
R_k(\alpha) = \frac{k}{2 \alpha} - 2\frac{\pi^2}{\alpha^2} L_{k-1}^{(1)}\left(\frac{2 \pi^2}{\alpha}\right) e^{-\frac{\pi^2}{\alpha}} + {\cal O}(e^{-4\frac{\pi^2}{\alpha}}/\alpha) \;.
\end{eqnarray}

We now want to analyze this formula for $R_k(\alpha)$ (\ref{rk.largedev.explicit}) when $k \sim {\cal O}(N)$ is large, see Eq. (\ref{start.expr.3}). For this purpose, we analyze the integral formula for $G_k(\xi)$ in Eq. (\ref{contour.gk}) using the saddle point method and write
\begin{eqnarray}\label{contour.gk.2}
G_k(\xi) =  \oint_{C_t} \frac{dt}{2\pi i} e^{-2 \xi t + k \ln{(1+1/t)}} \;, 
\end{eqnarray} 
which we analyze in the limit where $k, \xi \gg 1$ and $N \gg 1$ such that 
\begin{eqnarray}\label{def.y.T}
\frac{k}{N} = y \;, \; \frac{\xi}{N} = \frac{2L^2}{N} = T \;,
\end{eqnarray}
with $y$ and $T$ kept fixed. Therefore from Eq. (\ref{contour.gk.2}) one has
\begin{eqnarray}\label{def.S}
G_k(\xi) = \oint_{C_t} \frac{dt}{2\pi i} e^{-N S(t)} \;, \; S(t) = 2 T\,t - y \ln{\left(1 + \frac{1}{t}\right)} \;, 
\end{eqnarray}  
which can then be evaluated by a saddle point method when $N \to \infty$. This saddle point is such that $dS/dt|_{t=t^*} = 0$, where $S(t)$ reaches its maximum on the real axis, yielding
\begin{eqnarray}\label{saddle_point}
t^* = \frac{-1 + \sqrt{1-2y/T}}{2} \;.
\end{eqnarray}
It corresponds indeed to a maximum of $S(t)$ on the real line
\begin{eqnarray}\label{second.derivative}
S''(t^*) = - y \frac{2t^*+1}{(t^*(1+t^*))^2} < 0 \;.
\end{eqnarray}

Note that this saddle point solution (\ref{saddle_point}) makes sense if and only if $t^*$ is real, which implies
\begin{eqnarray}
2\frac{y}{T} < 1 \Rightarrow k < \frac{\xi}{2} = L^2 \;.
\end{eqnarray}
Now given that the largest value of $k$ involved in the expression of $\tilde F_L(N)$ in Eq. (\ref{start.expr.3}) is $k=2N+1$, this means that this saddle point analysis holds for
\begin{eqnarray}\label{limit.large.dev}
L > \sqrt{2N} \;,
\end{eqnarray}
which thus allows to study the right tail of the distribution. This yields finally 
\begin{eqnarray}\label{gk.saddle.eval}
G_k(\xi) \simeq -\frac{1}{\sqrt{2 \pi}} \frac{e^{-N S(t^*)}}{\sqrt{N |S''(t^*)|}} \;,
\end{eqnarray} 
where the minus sign in Eq. (\ref{gk.saddle.eval}) comes from the fact that the integral over the contour $C_t$ in Eq. (\ref{contour.gk.2}) is counterclockwise oriented, which thus runs, close to $t^*$ from $\Im(t) > 0$ to $\Im(t) < 0$ and hence the minus sign. One finally obtains, plugging the explicit expression of $t^*$ (\ref{saddle_point}) into Eq.~(\ref{gk.saddle.eval})
\begin{eqnarray}\label{gk.saddle.eval.2}
&&G_k(\xi) e^{-\xi} \simeq \frac{(-1)^{k-1}}{4} \frac{1}{\sqrt{\pi \xi}} \sqrt{\frac{2k}{\xi}} \left(1- \frac{2k}{\xi} \right)^{-\frac{1}{4}} \exp{\left(-\xi \gamma\left(\frac{2k}{\xi}\right)\right)} \;, \\
&&\gamma(x) = \sqrt{1-x} - \frac{x}{2} \ln{\left(\frac{1+\sqrt{1-x}}{1-\sqrt{1-x}} \right)} \;, \label{def.gamma}
\end{eqnarray}
where the prefactor $(-1)^k$ in Eq. (\ref{gk.saddle.eval.2}) comes the fact that $1+1/t^*<0$ such that when one evaluates $e^{-NS(t^*)}$ there is a $(-1)^k$ prefactor coming from $\exp{[k \ln(1+1/t^*)]}$ (\ref{def.S}). Note in particular the asymptotic behavior of $\gamma(x)$
\begin{eqnarray}\label{asympt_gamma}
\gamma(x) \sim \frac{2}{3}(1-x)^{\frac{3}{2}} \;, \; x \to 1 \;,
\end{eqnarray}
which will be useful in the following to understand the matching between large and typical fluctuations. Finally one has in this regime (\ref{def.y.T}):
\begin{eqnarray}\label{gk.saddle.rk}
R_k(\alpha) = \frac{k}{2\pi^2}\xi + (-1)^k \frac{\xi \sqrt{k}}{2^{1/2}\pi^{5/2}} \left(1 - \frac{2k}{\xi}\right)^{-1/4} \exp{\left[-\xi \gamma\left(\frac{2k}{\xi}\right)\right]} \;,
\end{eqnarray}
where we recall that $\xi = \pi^2/\alpha = 2L^2$. We now want to use this asymptotic behavior 
(\ref{gk.saddle.rk}) to study the right tail of the cumulative distribution $\tilde F_N(L)$. 
For this purpose, we start from the expression given in Eq. (\ref{omega.orthogonal.2}) to 
write (see Appendix A1)
\begin{eqnarray}\label{last_fn.1}
\tilde F_N(L) = \frac{\Gamma(N+1)}{\prod_{j=0}^{N-1} \Gamma(2+j)\Gamma(\frac{3}{2}+j)} \alpha^{N^2+\frac{N}{2}} \left[h_0(\alpha) R_1(\alpha)\right]^N \prod_{k=1}^{N-1} \left[R_{2k}(\alpha) R_{2k+1}(\alpha) \right]^{N-k} \;,
\end{eqnarray}
with $R_k(\alpha)$ given in Eq. (\ref{rk.ansatz.largedev}), which for $k \sim {\cal O}(N)$ reduces to Eq. (\ref{gk.saddle.rk}). To study this expression~(\ref{last_fn.1}) in the large deviation regime, we write $R_k(\alpha)$ as in Eq. (\ref{rk.ansatz.largedev}) and treat the term $c_k(\alpha) e^{-\frac{\pi^2}{\alpha}}$ in perturbation theory. After cumbersome manipulations left in Appendix \ref{app_large_dev} we arrive at the following formula
\begin{eqnarray}\label{last_fn.2}
\ln \tilde F_N(L) = e^{-\xi} \left[G_{2N}(\xi) + I_{2N}(\xi) \right] + {\cal O}\left(\exp{\left[-2\xi \gamma\left( \frac{2N}{\xi}\right) \right]} \right) \;,
\end{eqnarray}  
in terms of the variable $\xi$ given in Eq. (\ref{def_xi}) and the functions $G_{2N}(\xi)$ given in Eq. (\ref{gk.laguerre}) and $I_{2N}(\xi)$ given by
\begin{eqnarray}\label{expr_I2N}
I_{2N}(\xi) = - 2 \xi \sum_{k=0}^{N-1} \frac{G_{2k+1}(\xi)}{2k+1} \;.
\end{eqnarray}
This formula in Eq. (\ref{last_fn.2}) yields back the formula obtained in Ref.~\cite{CNS96} [see their Eq. (28)]. Using the integral representation for $G_k(\xi)$ obtained before in Eq. (\ref{contour.gk}) one can obtain an integral representation for $I_{2N}(\xi)$, see Appendix \ref{app_large_dev}, under the form
\begin{eqnarray}\label{i_integral}
I_{2N}(\xi) = \oint_{C_t} \frac{dt}{2\pi i} \frac{e^{-2 \xi t}}{1+2t} \left(1+\frac{1}{t} \right)^{2N} \;. 
\end{eqnarray}
The analysis of $I_{2N}(\xi)$ in the large deviation regime (\ref{def.y.T}) can then be carried out as before for $G_{2N}(\xi)$. By comparing the two formulas (\ref{contour.gk}) and (\ref{i_integral}) it is easy to see that in the large $L$ and large $N$ limit one has
\begin{eqnarray}\label{i.large.dev}
I_{2N}(\xi) \sim \frac{1}{1+2t^*} G_{2N}(\xi) = \frac{1}{\sqrt{1-4/T}}  G_{2N}(\xi) \;,
\end{eqnarray}
where we have used the expression for the saddle point in Eq. (\ref{saddle_point}). Finally one obtains that in the large deviation regime
\begin{eqnarray}\label{final.right.tail}
\ln \tilde F_N(L) \sim - \left[1 + \frac{1}{\sqrt{1-4N/\xi}} \right] \frac{1}{2\sqrt{\pi}} \frac{\sqrt{N}}{\xi} \left(1 - \frac{4N}{\xi}\right)^{-1/4} \exp{\left[-\xi \gamma\left(\frac{4N}{\xi}\right)\right]} \;,
\end{eqnarray}
which is negative, as it should, since $\tilde F_N(L) < 1$. Recalling that $\xi = 2L^2$ one obtains from Eq.~(\ref{final.right.tail}) the right tail behavior announced above in Eq. (\ref{main_results_F}),
\begin{eqnarray}
&&1-\tilde F_N(L) \simeq \exp{\left[-N \phi_A^+(L/\sqrt{2N})\right]} \;, \nonumber\\
&& \phi_A^+(x) = 4 x^2 \gamma(1/x^2) = 4 x \sqrt{x^2-1}-2 \ln{\left[2 x \left(\sqrt{x^2-1}+x\right)-1\right]} \label{explicit_phi+A} \;,
\end{eqnarray}
where we have used the explicit expression of $\gamma(x)$ in Eq. (\ref{gk.saddle.eval.2}). From the asymptotic behavior~(\ref{asympt_gamma}) one gets immediately
\begin{eqnarray}\label{asympt_rate_explicit}
\phi_A^+(x) = \frac{2^{9/2}}{3}(x-1)^{3/2} + {\cal O}\left((x-1)^{5/2}\right) \;, \; x \to 1^+ \;,
\end{eqnarray}
which yields the asymptotic behavior announced above in Eq. (\ref{asympt_rate}). Note in particular that if we set $L$ close to $\sqrt{2 N}$ such that
\begin{eqnarray}
L = \sqrt{2 N} + c N^{-\frac{1}{6}} s \;, \; c = 2^{-11/6} \;,
\end{eqnarray}
one has, using this asymptotic behavior in Eq. (\ref{asympt_gamma})
\begin{eqnarray}\label{large_dev_match}
\ln \tilde F_N(L) \sim - \frac{1}{4\sqrt{\pi} s^{3/4}}e^{-\frac{2}{3}s^{3/2}}  - N^{-1/3}\frac{1}{2^{8/3}\sqrt{\pi}s^{1/4}} e^{-\frac{2}{3}s^{3/2}} + {\cal O}(N^{-2/3})\;,
\end{eqnarray}
which is useful to study the matching, discussed in section 2, between the right tail and the central part of the distribution, which describes 
the typical fluctuations around $L = \sqrt{2N}$.

\subsection{Double scaling regime}\label{subsect.double.scaling}

We now focus on the typical fluctuations of the top path and study $\tilde F_N(L)$ for $L$ close to $\sqrt{2N}$. This corresponds to a double scaling limit where both $L$ and $N$ are large but such that $L - \sqrt{2N}= {\cal O}(N^{-1/6})$. Hence we set
\begin{eqnarray}\label{def_s}
L = \sqrt{2N} + c N^{-1/6} s \;, \; c = 2^{-11/6} \;.
\end{eqnarray}
We define a "running variable" $x_k$ as
\begin{eqnarray}\label{def_xk}
x_k = n_{cr}^{2/3}\left(1 - \frac{k}{n_{cr}} \right) \;, \; n_{cr} = \frac{\xi}{2} =  L^2 \;.
\end{eqnarray}
The goal is to analyze the recursion relations for $R_k(\alpha)$ in Eq. (\ref{volterra}) when $k$ is also of order ${\cal O}(N)$. This can be analyzed in this double scaling limit by assuming the following ansatz~\cite{GM94} (which was later on proved rigorously in Ref.~\cite{liechty})
\begin{eqnarray}\label{dble_scaling_ansatz}
&&R_k(\alpha) = \frac{n_{cr}^2}{\pi^2} + n_{cr}^{5/3} f_1^{+}(x_k) + n_{cr}^{4/3} f_2^{+}(x_k) + n_{cr} f_3^{+}(x_k) + {\cal O}(n_{cr}^{2/3}) \;, \; k \; {\rm even }\;, \\
&&R_k(\alpha) = \frac{n_{cr}^2}{\pi^2} + n_{cr}^{5/3} f_1^{-}(x_k) + n_{cr}^{4/3} f_2^{-}(x_k) + n_{cr} f_3^{-}(x_k) + {\cal O}(n_{cr}^{2/3}) \;, \; k \; {\rm odd }\;. \nonumber 
\end{eqnarray} 
The fact that we assume different functions $f_1^{\pm}, f_2^{\pm}, f_3^{\pm}$ depending on the parity of $k$ is guided by the previous result (\ref{gk.saddle.rk}) where we showed that the leading correction comes with an amplitude proportional to $(-1)^k$, hence depending on the parity of $k$. By substituting this ansatz (\ref{dble_scaling_ansatz}) into the recursion relation in Eq. (\ref{volterra}) one obtains that 
\begin{eqnarray}\label{f1}
f_1^+(x) = - f_1^-(x) = -f_1(x) \;, 
\end{eqnarray}
where the function $f_1$ satisfies a Painlev\'e II equation (PII)
\begin{eqnarray}\label{eq_f1}
f_1''(x) = 4 x f_1(x) + \frac{\pi^4}{2} f_1^3(x) \;, \; 
f_1(x) \mathop{\sim}\limits_{x\rightarrow \infty}  - \frac{2^{5/3}}{\pi^2}{\rm Ai}(2^{2/3} x) \;,
\end{eqnarray}
where ${\rm Ai}(x)$ is the standard Airy function. Note the minus sign in the asymptotic 
behavior which is missing in Ref. \cite{GM94}. It can be expressed in terms of the 
Hastings-McLeod solution of PII $q(s)$ in Eq.~(\ref{PII}) as

\begin{eqnarray}\label{rel_f1_mcleod}
f_1(x) = -\frac{2^{5/3}}{\pi^2} q(2^{2/3} x) \;,
\end{eqnarray} 
where
\begin{equation}
q''(s) = 2q^3(s) + s q(s) \;, \; q(s)\mathop{\sim}\limits_{s\rightarrow \infty} {\rm Ai}(s) \mathop{\sim}\limits_{s\rightarrow \infty} \frac{1}{2\sqrt{\pi}s^{1/4}} e^{-\frac{2}{3}s^{3/2}}   \;.
\label{qdiff}
\end{equation}
One can further show \cite{GM94} that the functions $f_2^+, f_2^-$ in Eq. (\ref{dble_scaling_ansatz}) satisfy
\begin{eqnarray}\label{f2}
f_2^+(x) + f_2^-(x) = - \frac{2}{\pi^2} x + \frac{\pi^2}{2} f_1^2(x) \;,
\end{eqnarray}
and that the functions $f_3^+, f_3^-$ in Eq. (\ref{dble_scaling_ansatz}) satisfy
\begin{eqnarray}
f_3^+(x) + f_3^-(x) = \frac{1}{3} x f_1(x) + \pi^2 f_1(x) f_2^-(x) - \frac{\pi^4}{3} f_1^3(x) + \frac{1}{6} f_1''(x) \;.
\end{eqnarray}
Using this ansatz (\ref{dble_scaling_ansatz}), one can then analyze $\tilde F_N(L)$ in the double scaling limit. For this purpose, it is useful to start from Eq. (\ref{start.expr.3}): taking the logarithm on both sides and performing the large $n_{cr}$ expansion of the right hand side, one obtains after a tedious though straightforward calculation, using (\ref{f1}, \ref{f2}):
\begin{eqnarray}\label{exp_rhs}
\ln \tilde F(L,N+1) + \ln \tilde F_{N-1}(L) &-& 2 \ln \tilde F_{N}(L) = -n_{cr}^{-2/3}\frac{\pi^4}{2} \left(f_1^2(x_{2N})  + \frac{2}{\pi^2}f_1'(x_{2N})  \right) \\
&+& n_{cr}^{-1} \frac{\pi^4}{2} \left(f_1(x_{2N})f_1'(x_{2N}) + \frac{1}{\pi^2}f_1''(x_{2N}) \right) + {\cal O}(n_{cr}^{-4/3}) \nonumber \;.
\end{eqnarray} 
On the other hand, if we assume the following ansatz for $\ln \tilde F_N(L)$
\begin{eqnarray}\label{ansatz_fn}
\ln \tilde F_N(L) = Y(x_{2N}) + n_{cr}^{-1/3} H(x_{2N}) + {\cal O}(n_{cr}^{-2/3})\;,
\end{eqnarray}
the expansion of the left hand side of Eq. (\ref{exp_rhs}) yields
\begin{eqnarray}\label{exp_lhs}
\ln \tilde F_{N+1}(L) + \ln \tilde F_{N-1}(L) - 2 \ln \tilde F_{N}(L) = 4 n_{cr}^{-2/3} Y''(x_{2N}) + 4 n_{cr}^{-1} H''(x_{2N}) + {\cal O}(n_{cr}^{-4/3}) \;.
\end{eqnarray}
Therefore identifying the terms with the same power of $n_{cr}$ in Eqs. (\ref{exp_rhs}) and (\ref{exp_lhs}) one obtains
\begin{eqnarray}\label{relation_yh}
&&4 Y''(x) =  -\frac{\pi^4}{2} \left(f_1(x_{})^2  +\frac{2}{\pi^2}f_1'(x_{})  \right) \\
&&4 H''(x) = \frac{\pi^4}{2} \left(f_1(x_{})f_1'(x_{}) + \frac{1}{\pi^2}f_1''(x_{}) \right) \;. \nonumber
\end{eqnarray}
In terms of the variable $s$ defined in Eq. (\ref{def_s}) one has from (\ref{def_xk})
\begin{eqnarray}\label{rel_xs_1}
x_{2N} = 2^{-2/3} s \;,
\end{eqnarray}
and if we define
\begin{eqnarray}\label{def_ytilde}
Y(x) =\tilde Y(2^{2/3} x) \;, \; H(x) = \tilde H(2^{2/3}x) \;,
\end{eqnarray}
the above equations (\ref{relation_yh}) can then be written in terms of $q(s)$, the Hastings-McLeod solution of PII (\ref{PII}) as
\begin{eqnarray}\label{yh_seconde}
&&\tilde Y''(s)  = -\frac{1}{2} \left(q^2(s) - q'(s) \right) \\
&&\tilde H''(s) = 2^{-1/3} \left(q(s) q'(s) - \frac{1}{2}q''(s) \right) = -2^{-1/3} Y'''(s) \;. \nonumber
\end{eqnarray}
Therefore, from Eqs (\ref{ansatz_fn}), (\ref{def_ytilde}) together with Eq. (\ref{yh_seconde}) one obtains, using that $\tilde F_N(L \to \infty) = 1$
\begin{eqnarray}
&&\ln \tilde F_N(\sqrt{2N} + 2^{-11/6} N^{-1/6} s) = \tilde Y(s) - N^{-1/3} 2^{-2/3} \tilde Y'(s) + {\cal O}(N^{-2/3}) \\
&& Y(s) = - \frac{1}{2} \int_{s}^\infty (x-s) q^2(x) \, dx - \frac{1}{2} \int_s^\infty q(x) \, dx \;.
\end{eqnarray}
So that finally one has
\begin{eqnarray}\label{final_fN}
\tilde F_N(\sqrt{2N} + 2^{-11/6} N^{-1/6} s) = {\cal F}_1(s) - N^{-1/3} 2^{-2/3} {\cal F}_1'(s) + {\cal O}(N^{-2/3}) \;.
\end{eqnarray}
Note that using the large $s$ behavior of $q(s)$ in Eq. (\ref{qdiff}) one obtains the large $s$ behavior of ${\cal F}_1(s)$ as:
\begin{eqnarray}
\ln {\cal F}_1(s) \sim - \frac{1}{2} \int_s^\infty q(x) \, dx \sim - \frac{1}{4\sqrt{\pi} s^{3/4}} e^{-\frac{2}{3}s^{3/2}} (1 + {\cal O}(s^{-3/2})) \;, 
\end{eqnarray}
and therefore, expanding each term of the large $N$ expansion in Eq.~(\ref{final_fN}) for large $s$, one gets
\begin{eqnarray}
\ln \tilde F_N(\sqrt{2N} + 2^{-11/6} N^{-1/6} s) = &-& \frac{1}{4\sqrt{\pi} s^{3/4}} e^{-\frac{2}{3}s^{3/2}}(1 + {\cal O}(s^{-3/2}))  \\
&-& N^{-1/3} \frac{1}{2^{8/3}\sqrt{\pi} s^{1/4}} e^{-\frac{2}{3}s^{3/2}}(1 + {\cal O}(s^{-3/2}))  +  {\cal O}(N^{-2/3}) \;. \nonumber
\end{eqnarray}
This expansion matches perfectly with the large deviation behavior obtained above in Eq. (\ref{large_dev_match}). We notice that this is not only true for the leading term of order ${\cal O}(N^0)$ but also for the subleading one, of order ${\cal O}(N^{-1/3})$. 

Although the leading term in Eq. (\ref{final_fN}) was already obtained in Ref. \cite{FMS11}, and subsequently shown rigorously in \cite{liechty}, we also obtain here the first correction, of order ${\cal O}(N^{-1/3})$ to ${\cal F}_1$. Note that this correction, being proportional to ${\cal F}_1'(s)$, can also be written as a simple shift of the argument of ${\cal F}_1$ as
\begin{eqnarray}\label{shift_F}
\tilde F_N(\sqrt{2N} + 2^{-11/6} N^{-1/6} s) = {\cal F}_1(s - N^{-1/3} 2^{-2/3}) + {\cal O}(N^{-2/3}) \;,
\end{eqnarray}
such that this leading correction only affects the first moment, the finite $N$ corrections to the higher cumulants being at least of order ${\cal O}(N^{-2/3})$ (see also Ref. \cite{baik_jenkins} for a related computation in a sightly different context).

\section{Large $N$ analysis of the reunion probability of $N$ non-intersecting Brownian motions on a circle}\label{largeN_B}

In this section we focus on the ratio between reunion probabilities $\tilde G_N(L)$ for non-intersecting Brownian motions on a circle, i.e. with
periodic boundary conditions. We start with the expression given in section 2 in Eq.~(\ref{29})
\begin{eqnarray}\label{start_expr_g}
\tilde G_N(L) = \frac{B_N}{L^{N^2}} \sum_{n_1 = -\infty}^\infty \cdots 
\sum_{n_N=-\infty}^\infty \Delta_N^2(n_1, \cdots,n_N)\, e^{-\frac{2 \pi^2}{L^2} \sum_{j=1}^N 
n_j^2} 
\end{eqnarray}
where $\Delta_N(y_1, \cdots, y_N)$ is the Vandermonde determinant (\ref{vdm.det}) and the prefactor $B_N$ is given by
\begin{eqnarray}\label{AN}
B_N = \frac{1}{(2 \pi)^{N/2-N^2} \prod_{j=0}^{N-1} \Gamma(j+2)} \;.
\end{eqnarray} 
Here we introduce the parameter $\alpha$ defined as
\begin{eqnarray}\label{alpha_periodic}
\alpha = \frac{2 \pi^2}{L^2} \;,
\end{eqnarray}
and rewrite
\begin{equation}\label{start_expr_g1}
\tilde G_N(L) = 
\frac{\alpha^{N^2/2}}{\prod_{j=0}^{N-1} 
\Gamma(j+2) 2^{N/2(1-N)} \pi^{N/2}} 
\sum_{n_1 = -\infty}^\infty \cdots 
\sum_{n_N=-\infty}^\infty \Delta_N(n_1, \cdots,n_N)^2 e^{-\alpha \sum_{j=1}^N n_j^2} \;.
\end{equation}
Note that, for simplicity, we use the same notation $\alpha$ as in model A, and in model C 
below, although its relation to $L$ differs from one model (\ref{alpha_absorbing}), 
(\ref{alpha_reflecting}) to another (\ref{alpha_periodic}).

\subsection{Coulomb gas analysis for large $N$ and the left tail of $\tilde G_N(L)$}

This case corresponds exactly, up to a multiplicative prefactor, to the partition function of Yang-Mills theory on the sphere with the gauge group ${\rm U}(N)$, which was analyzed by Douglas and Kazakov in Ref. \cite{DK93}. The transposition of their results to the present study was already presented in Ref. \cite{FMS11}. There it was shown that the transition for $\tilde G_N(L)$ happens for $L = 2 \sqrt{N}$ such that, if one introduces the parameter $r = L/2\sqrt{N}$ one has, similarly to what we have obtained before for model A (\ref{main_res_CG}) 
\begin{eqnarray}\label{main_res_CG_periodic}
\lim_{N \to \infty} -\frac{1}{N^2} \ln G_N(\sqrt{4N} \, r) = 
\begin{cases}
0 \;, \; r \geq 1 \\
\phi_B^-(r)=\frac{1}{2}\phi_A^-(r)  \;, \; r < 1 \;,
\end{cases}
\end{eqnarray} 
where the function $\phi_A^-$ is defined in Eq. (\ref{main_res_CG}). Here, the asymptotic behavior of $\tilde G_N(2 \sqrt{N} \, r)$ for $r$ close to $1$ is thus
\begin{eqnarray}\label{left_tail_asympt}
\tilde G_N(\sqrt{4N} r) \sim \exp{\left(-\frac{8}{3}N^2 (1-r)^{3} \right)} \;.
\end{eqnarray}
These results in Eq.~(\ref{main_res_CG_periodic}) and Eq.~(\ref{left_tail_asympt}) yield the behaviors announced in section 2 in Eqs (\ref{main_results_G}), (\ref{asympt_rate_B}), for the left tail of $\tilde G_N(L)$. However, as before for model A, this saddle point method does not give anything meaningful for the right tail, which can be analyzed using the method of discrete orthogonal polynomials described before. 

\subsection{Analysis of the right tail of $\tilde G_N(L)$}

The right hand side of Eq. (\ref{start_expr_g1}) can be conveniently expressed in terms of 
the orthogonal polynomials 
$p_k(n)$ introduced and studied above (\ref{ortho_condition}) where $\alpha$ is now defined by (\ref{alpha_periodic}). Using standard manipulations (in particular the Cauchy-Binet formula), one arrives at
\begin{eqnarray}\label{expr_g_ortho}
\tilde G_N(L) =  \frac{\alpha^{N^2/2} \Gamma(N+1)}{\prod_{j=0}^{N-1} \Gamma(j+2) 2^{N/2(1-N)} \pi^{N/2}} \prod_{j=0}^{N-1} h_j(\alpha) \;.
\end{eqnarray}
As we have done before (\ref{last_fn.1}), this product of the amplitudes $h_j$ in Eq. 
(\ref{expr_g_ortho}) can be re-written in terms of the $R_k$'s. Substituting then the expression for $R_k$'s in Eq. (\ref{Rk}) and treating the terms $c_k(\alpha) e^{-\frac{\pi^2}{\alpha}}$ in perturbation theory, one obtains (see also Ref. \cite{GM94})
\begin{eqnarray}
\ln \tilde G_N(L) = 2 e^{-\xi} G_{N}(\xi)  + {\cal O}\left(\exp{\left[-2\xi \gamma\left( \frac{2N}{\xi}\right) \right]} \right)\;,
\end{eqnarray}
in terms of the variable $\xi = \pi^2/\alpha = L^2/2$, reminding that $\alpha = 2\pi^2/L^2$ in this case (\ref{alpha_periodic}). This yields, using the expression for $G_N(\xi)$ given in Eq. (\ref{gk.saddle.eval.2}):
\begin{eqnarray}\label{expr_g_largedev}
\ln \tilde G_N(L) &=& \frac{(-1)^{N-1}}{2} \frac{1}{\sqrt{\pi \xi}} \sqrt{\frac{2N}{\xi}} \left(1 - \frac{2N}{\xi} \right)^{-1/4} \exp{\left[-\xi \gamma\left( \frac{2N}{\xi}\right) \right]} \\
&=&\frac{(-1)^{N-1}}{\sqrt{2 \pi} L} \sqrt{\frac{4N}{L^2}} \left(1 - \frac{4N}{L^2} \right)^{-1/4} \exp{\left[-\frac{L^2}{2} \gamma\left(\frac{4N}{L^2} \right) \right]} \;,
\end{eqnarray}
which shows an interesting oscillatory behavior with $N$: this fact is not problematic as $\tilde G_N(L)$ does not have the meaning of a cumulative distribution. One thus obtains from Eq. (\ref{expr_g_largedev}):
\begin{eqnarray}
1-\tilde G_N(L) = (-1)^N \exp{\left[- N \phi_B^+(L/\sqrt{4N})\right]} \;, \; \phi_B^+(x) = 2 x^2 \gamma\left(\frac{1}{x^2}\right) = \frac{1}{2}\phi_A^+(x) \:,
\end{eqnarray}
as announced earlier in section 2 (\ref{main_results_G}), where we have used the expression of $\phi_A^+(x)$ given in Eq.~(\ref{explicit_phi+A}). The asymptotic behavior of $\phi_B^+(x)$ when $x \to 1$ follows immediately from Eq. (\ref{asympt_rate_explicit}), as announced in Eq. (\ref{asympt_rate_B}). Therefore, if we study the regime of $L$ close to $2 \sqrt{N}$ and set
\begin{eqnarray}
L = 2 \sqrt{N} + 2^{-2/3} N^{-1/6} \, t \;,
\end{eqnarray}
one obtains from Eq. (\ref{expr_g_largedev})
\begin{eqnarray}\label{largedev_matching}
\ln \tilde G_N(L) \sim (-1)^{N-1} N^{-1/3} \frac{1}{2^{4/3} \sqrt{\pi} t^{1/4}} e^{-\frac{2}{3}t^{3/2}} \;, 
\end{eqnarray}
where we notice that this term is of order ${\cal O}(N^{-1/3})$: this is responsible for the (mis-)matching with the central part of the distribution, which is instead of order ${\cal O}(N^0)$, and described by a double scaling limit which we now focus on.

\subsection{Double scaling regime: analysis of the central part of the distribution}

To analyze the central part of $\tilde G_N(L)$ we start with the following identity, analogous to the one in Eq. (\ref{start.expr.3}) for model A, which reads here
\begin{eqnarray}\label{start_expr_dble_g}
\frac{\tilde G_{N+1}(L) \tilde G_{N-1}(L)}{[\tilde G_N(L)]^2} = \frac{2\alpha}{N} \frac{h_N}{h_{N-1}} = \frac{2\alpha}{N} R_N(\alpha) \;,
\end{eqnarray}
which is very useful for asymptotic analysis. Using the asymptotic expansion of $R_N(\alpha)$ in Eq.~(\ref{dble_scaling_ansatz}), one can expand the right hand side of Eq.~(\ref{start_expr_dble_g}) as
\begin{eqnarray}\label{expansion_rhs_G}
\ln{\left( \frac{2\alpha}{N} R_N(\alpha) \right)}= (-1)^{N-1} \pi^2 f_1(x_N) n_{cr}^{-1/3} - \frac{\pi^4}{4} f_1(x_N)^2 n_{cr}^{-2/3} + {\cal O}(n_{cr}^{-1}) \;, 
\end{eqnarray}  
where $n_{cr}$ is given here by
\begin{eqnarray}
n_{cr} = \frac{L^2}{4} \;.
\end{eqnarray}
On the other hand, guided by the result which we have obtained for the right tail of $\tilde G_N(L)$ and also by the expansion above (\ref{expansion_rhs_G}), one assumes the following ansatz for $\ln \tilde G_N(L)$ in the double scaling regime
\begin{eqnarray}
\ln \tilde G_N(L) = Y(x_N) + (-1)^{N-1} n_{cr}^{-1/3} H(x_N) \;, 
\end{eqnarray}
such that the left hand side of Eq. (\ref{start_expr_dble_g}) admits the following expansion
\begin{equation}\label{expansion_lhs_G}
\ln \tilde G_{N+1}(L)+\ln \tilde G_{N-1}(L)-2\ln \tilde G_N(L) = (-1)^N n_{cr}^{-1/3} 4 H(x_N) + n_{cr}^{-2/3} Y''(x_N) + {\cal O}(n_{cr}^{-1}) \;.
\end{equation}
Therefore, identifying the different powers of $n_{cr}$ in Eqs. (\ref{expansion_rhs_G}) and (\ref{expansion_lhs_G}) one finds
\begin{eqnarray}\label{identification_G}
Y''(x) = - \frac{\pi^4}{4} f_1(x)^2 \;,\; H(x) =  -\frac{\pi^2}{4} f_1(x) \;.
\end{eqnarray}
If we set 
\begin{eqnarray}
Y(x) = \tilde Y(2^{2/3} x) \;, \; H(x) = \tilde H(2^{2/3}x) \;,
\end{eqnarray}
one has from Eq. (\ref{identification_G})
\begin{eqnarray}
\tilde Y''(s) = - q^2(s) \;, \; \tilde H(s) = 2^{-1/3} q(s) \;.
\end{eqnarray}
One thus obtains
\begin{eqnarray}
&&\ln \tilde G_N(2 \sqrt{N} + 2^{-2/3} s N^{-1/6}) = \tilde Y(s) + (-1)^{N-1} N^{-1/3} 2^{-1/3} q(s) + {\cal O}(N^{-2/3})\\
&& \tilde Y(s) = - \int_{s}^\infty (x-s) q^2(x) \, dx \;,
\end{eqnarray}
which yields
\begin{eqnarray}
\lim_{N \to \infty} \tilde G_N(2 \sqrt{N} + 2^{-2/3} s N^{-1/6}) = {\cal F}_2(s) \:.
\end{eqnarray}
The result in Eq.~(\ref{expansion_lhs_G}) yields the formula given in Eq. (\ref{G_crossover}) in section 2. At variance with model A (and model C discussed below) the competition between the two terms ${\cal F}_2(s) \sim \exp{(-\frac{4}{3}s^{3/2})}$ and $q(s) \sim \exp{(-\frac{2}{3}s^{3/2})}$ yields an unconventional crossover regime discussed in section 2 and illustrated in Fig. \ref{fig_crossover}, which is peculiar to this vicious walker problem.

\section{Large $N$ analysis of the reunion probability of $N$ non-intersecting Brownian motions with reflecting boundary conditions}\label{largeN_C}

In this section we focus on the ratio of reunion probabilities $\tilde E_N(L)$ for non-intersecting Brownian motions on the segment $[0,L]$ with reflecting boundary conditions at $x=0$ and $x=L$. We start with the expression given in section 2 in Eq.~(\ref{ENL_intro})
\begin{equation}\label{ENL}
\tilde{E}_N(L)= \frac{C_N}{L^{2N^2-N}}\sum_{n_1=-\infty}^\infty
\ldots \sum_{n_N = -\infty}^{\infty}
\Delta^2(n_1^2,\ldots,n_N^2) e^{-\frac{\pi^2}{2 L^2} \sum_{j=1}^N n_j^2} \;,
\end{equation}
where $\Delta_N(y_1, \cdots, y_N)$ is the Vandermonde determinant (\ref{vdm.det})
and the prefactor $C_N$ given by
\begin{equation}
C_N=\frac{{\pi}^{2N^2-N}\,2^{N/2-N^2}}{\prod_{j=0}^{N-1}\Gamma(2+j)\Gamma(1/2+j)} \;,
\label{norm-cn1}
\end{equation}
ensures that $\tilde{E}_N(L\to \infty)=1$. Here we introduce the parameter $\alpha$ given, as in model A (\ref{alpha_absorbing}), by
\begin{eqnarray}\label{alpha_reflecting}
\alpha = \frac{\pi^2}{2L^2} \;,
\end{eqnarray} 
and write
\begin{equation}\label{start_expr_e1}
\tilde E_N(L) =  \frac{\alpha^{N^2-\frac{N}{2}}}{\prod_{j=0}^{N-1}\Gamma(2+j)\Gamma(1/2+j)} \sum_{n_1=-\infty}^\infty
\ldots \sum_{n_N = -\infty}^{\infty}
\Delta^2(n_1^2,\ldots,n_N^2) e^{-\alpha \sum_{j=1}^N n_j^2} \;.
\end{equation}

\subsection{Coulomb gas analysis for large $N$ and the left tail of $\tilde E_N(L)$}

The Coulomb gas analysis of $\tilde E_N(L)$, which amounts to the study of the path-integral (over the density $\tilde \rho$) entering the expression of $\tilde E_N(L)$ as in Eq. (\ref{expr_cg}), is exactly the same as the one done for $\tilde F_N(L)$. Indeed the difference between $\tilde F_N(L)$ and $\tilde E_N(L)$ is the presence, in the expression for $\tilde F_N(L)$, of the product $\prod_{i=1}^N n_i^2$. But this term does not contribute to $S[\tilde \rho]$ as $\ln \prod_{i=1}^N n_i^2 \sim {\cal O}(N)$, and are thus subdominant compared to ${\cal O}(N^2)$ terms which contribute to $S[\tilde \rho]$. Therefore one gets immediately the result for the left tail of  $\tilde E_N(L)$
\begin{eqnarray}\label{main_res_CG_E}
\lim_{N \to \infty} -\frac{1}{N^2} \ln \tilde E_N(h\sqrt{2N}) = 
\begin{cases}
0 \;, \; h \geq 1 \\
\phi_C^-(h) = \phi_A^-(h) \;, \; h < 1 \;,
\end{cases}
\end{eqnarray} 
where the rate function $\phi_A^-(h)$ is given in Eq. (\ref{main_res_CG}). This yields the result for the left tail of $\tilde E_N(L)$ announced in Ref. (\ref{main_results_E}).

\subsection{Analysis of the right tail of $\tilde E_N(L)$}

The expression above (\ref{start_expr_e1}) can be conveniently expressed in terms of the orthogonal polynomials $p_k(n)$ introduced and studied above (\ref{ortho_condition}) where $\alpha = \pi^2/2L^2$ (\ref{alpha_reflecting}). Using the identity
\begin{eqnarray}
\Delta_N(n_1^2, \cdots, n_N^2) = \det_{1 \leq i,j \leq N} \left[p_{2i-2}(n_j)\right] \;,
\end{eqnarray}
and the Cauchy-Binet formula, one arrives at
\begin{eqnarray}\label{expr_g_ortho_c}
\tilde E_N(L) =    \frac{\alpha^{N^2-\frac{N}{2}}}{\prod_{j=1}^{N-1}\Gamma(2+j)\Gamma(1/2+j)} 
\prod_{j=1}^{N} h_{2j-2}(\alpha) \;.
\end{eqnarray}
As we have done before (\ref{last_fn.1}), this product of the amplitudes $h_j$ in Eq. 
(\ref{expr_g_ortho_c}) can be re-written in terms of the $R_k$'s. Substituting then the 
expression for $R_k$'s in Eq. (\ref{Rk}) and treating the terms $c_k(\alpha) e^{-\frac{\pi^2}{\alpha}}$ in perturbation theory, one obtains [see also Ref. \cite{CNS96}]
\begin{eqnarray}
\ln \tilde E_N(L) = e^{-\xi} \left [G_{N}(\xi) - I_{2N}(\xi) \right]  + {\cal O}\left(\exp{\left[-2\xi \gamma\left( \frac{2N}{\xi}\right) \right]} \right)\;,
\end{eqnarray}
in terms of the variable $\xi = \pi^2/\alpha = 2 L^2$, where we have used $\alpha = \pi^2/2L^2$ (\ref{alpha_reflecting}). Note that this expression is very similar to, albeit different from, the expression obtained for $\ln \tilde F_N(L)$ in Eq.~(\ref{last_fn.2}) (the term $I_{2N}(\xi)$ comes here with a minus sign). One thus gets, from the previous analysis~(\ref{final.right.tail}), 
\begin{eqnarray}\label{final.right.tail.E}
\ln \tilde E_N(L) \sim - \left[1 - 
\frac{1}{\sqrt{1-4N/\xi}} \right] \frac{1}{2\sqrt{\pi}} 
\frac{\sqrt{N}}{\xi} \left(1 - \frac{4N}{\xi}\right)^{-1/4} 
\exp{\left[-\xi \gamma\left(\frac{4N}{\xi}\right)\right]}\; .
\end{eqnarray}
Therefore, one obtains from Eq. 
(\ref{final.right.tail.E})
\begin{eqnarray}
1 - \tilde E_N(L) = - \exp{[-N \phi_C^+(L/\sqrt{2N})]} \;, \; \phi_C^+(x) = 4 x^2 \gamma \left(\frac{1}{x^2}\right) = \phi_A^+(x) \;, 
\end{eqnarray}
as announced in Eq. (\ref{main_results_E}). The asymptotic behavior of $ \phi_C^+(x)$ when $x \to 1$ follows straightforwardly from Eq. (\ref{asympt_rate_explicit}). Note in particular that if we set $L$ close to $\sqrt{2 N}$ such that
\begin{eqnarray}\label{def_s_C}
L = \sqrt{2 N} + c N^{-\frac{1}{6}} s \;, \; c = 2^{-11/6} \;,
\end{eqnarray}
one has, from Eq. (\ref{final.right.tail.E}), using the asymptotic behavior for $\gamma(x)$ in Eq. (\ref{asympt_gamma})
\begin{eqnarray}\label{large_dev_match_C}
\ln \tilde E_N(L) \sim \frac{1}{4\sqrt{\pi} s^{3/4}}e^{-\frac{2}{3}s^{3/2}}  - N^{-1/3}\frac{1}{2^{8/3}\sqrt{\pi}s^{1/4}} e^{-\frac{2}{3}s^{3/2}} + {\cal O}(N^{-2/3})\;,
\end{eqnarray}
which is useful to study the matching with the typical fluctuations around $L = \sqrt{2N}$.

\subsection{Double scaling regime: analysis of the central part of the distribution}

To analyze the central part of $\tilde E_N(L)$ we start with the following identity, analogous to the one in Eq. (\ref{start.expr.3}) for model A, which reads here
\begin{eqnarray}\label{start_expr_dble_e}
\frac{\tilde E_{N+1}(L) \tilde E_{N-1}(L)}{[\tilde E_N(L)]^2} = \frac{\alpha^2}{N(N-1/2)} \frac{h_{2N}}{h_{2N-2}} = \frac{\alpha^2}{N(N-1/2)} R_{2N}(\alpha) R_{2N-1}(\alpha)\;, 
\end{eqnarray}
which is very useful for asymptotic analysis. Using the asymptotic expansion of $R_N(\alpha)$ in Eq.~(\ref{dble_scaling_ansatz}), one can expand the right hand side of Eq.~(\ref{start_expr_dble_e}) as
\begin{eqnarray}\label{exp_rhs_C}
\ln \left(\frac{\alpha^2}{N(N-1/2)} R_{2N}(\alpha) R_{2N-1}(\alpha) \right) &=& -n_{cr}^{-2/3}\frac{\pi^4}{2} \left(f_1^2(x_{2N})  - \frac{2}{\pi^2}f_1'(x_{2N})  \right) \\
&+& n_{cr}^{-1} \frac{\pi^4}{2} \left(-f_1(x_{2N})f_1'(x_{2N}) + \frac{1}{\pi^2}f_1''(x_{2N}) \right) + {\cal O}(n_{cr}^{-4/3}) \nonumber \;,
\end{eqnarray}
where, here
\begin{eqnarray}
n_{cr} = L^2 \;.
\end{eqnarray}
On the other hand, if we assume the following ansatz for $\ln \tilde E_N(L)$
\begin{eqnarray}\label{ansatz_en}
\ln \tilde E_N(L) = Y(x_{2N}) + n_{cr}^{-1/3} H(x_{2N}) + {\cal O}(n_{cr}^{-2/3})\;,
\end{eqnarray}
the expansion of the left hand side of Eq. (\ref{exp_rhs_C}) reads
\begin{equation}\label{exp_lhs_C}
\ln \tilde E_{N+1}(L) + \ln \tilde E_{N-1}(L) - 2 \ln \tilde E_{N}(L) = 4 n_{cr}^{-2/3} Y''(x_{2N}) + 4 n_{cr}^{-1} H''(x_{2N}) + {\cal O}(n_{cr}^{-4/3}) \;.
\end{equation}
Therefore identifying the terms in Eqs. (\ref{exp_rhs_C}) and Eq. (\ref{exp_lhs_C}) with the same power of $n_{cr}$ one obtains
\begin{eqnarray}\label{relation_yh_C}
&&4 Y''(x) =  -\frac{\pi^4}{2} \left(f_1(x_{})^2  -\frac{2}{\pi^2}f_1'(x_{})  \right) \\
&&4 H''(x) = -\frac{\pi^4}{2} \left(f_1(x_{})f_1'(x_{}) - \frac{1}{\pi^2}f_1''(x_{}) \right) \;. \nonumber
\end{eqnarray}
In terms of the variable $s$ defined in Eq. (\ref{def_s_C}) one has from (\ref{def_xk})
\begin{eqnarray}\label{rel_xs}
x_{2N} = 2^{-2/3} s \;,
\end{eqnarray}
and if we define
\begin{eqnarray}\label{def_ytilde_C}
Y(x) =\tilde Y(2^{2/3} x) \;,  H(x) = \tilde H(2^{2/3}x) \;,
\end{eqnarray}
the above equations (\ref{relation_yh_C}) can then be written in terms of $q(s)$, the Hastings-McLeod solution of PII (\ref{PII}) as
\begin{eqnarray}\label{yh_seconde_C}
&&\tilde Y''(s)  = -\frac{1}{2} \left(q^2(s) + q'(s) \right) \\
&&\tilde H''(s) = 2^{-1/3} \left(q(s) q'(s) + \frac{1}{2}q''(s) \right) = 2^{-1/3} \tilde Y'''(s) \;. \nonumber
\end{eqnarray}
Therefore, from Eqs. (\ref{ansatz_en}) and (\ref{def_ytilde_C}) together with Eq. (\ref{yh_seconde_C}) one obtains, using that $\tilde E_N(L \to \infty) = 1$
\begin{eqnarray}
&&\ln \tilde E_N(\sqrt{2N} + 2^{-11/6} N^{-1/6} s) = \tilde Y(s) + N^{-1/3} 2^{-2/3} \tilde Y'(s) + {\cal O}(N^{-2/3}) \\
&& Y(s) = - \frac{1}{2} \int_{s}^\infty (x-s) q^2(x) \, dx + \frac{1}{2} \int_s^\infty q(x) \, dx \;.
\end{eqnarray}
So that finally one has
\begin{eqnarray}\label{final_EN}
\tilde E_N(\sqrt{2N} + 2^{-11/6} N^{-1/6} s) = \frac{{\cal F}_2(s)}{{\cal F}_1(s)} + N^{-1/3} 2^{-2/3} \frac{d}{ds} \left[ \frac{{\cal F}_2(s)}{{\cal F}_1(s)} \right] 
+ {\cal O}(N^{-2/3}) \;,
\end{eqnarray}
where we have used the expression of ${\cal F}_2$ and ${\cal F}_1$ in terms of $q(s)$ given respectively in Eq. (\ref{F2}) and (\ref{F2-GOE}). As noticed previously for model A (\ref{shift_F}), the first correction in Eq. (\ref{final_EN}), proportional to $N^{-1/3}$, can be absorbed by a shift of the argument $s \to s + 2^{-2/3}N^{-1/3}$. An analogous structure is known for the large $N$ expansion at the soft edge
of the Laguerre ensemble in random matrix theory (see \cite{Fo10}, eq. (7.162)).

\section{Conclusion}

To conclude, we have performed in this paper a systematic study of three different models 
of $N$ non-intersecting Brownian motions on a line segment $[0,L]$ with three different types 
of boundary conditions at $x=0$ and $x=L$: absorbing (model A), periodic (model B) and 
reflecting (model C) boundary conditions. In each of these models we have focused on a 
normalized reunion probability which, in model A, can also be interpreted as the maximal 
height of $N$ non-intersecting Brownian excursions on the unit time interval. We have 
presented a self-contained derivation of the formulas for finite $N$ for these reunion 
probabilities, whose expressions had been given without the details in previous publications 
in Ref. \cite{SMCR08}, for model A, and for model B and C in Ref. \cite{FMS11}. An 
interesting property of these reunion probabilities is that they are, up to a multiplicative 
pre-factor, identical to the partition function of Yang-Mills theory on the sphere with a 
gauge group $G$ which is selected by the choice of boundary conditions: ${\rm Sp}(2N)$ for 
model~A, ${\rm U}(N)$ for model B and ${\rm SO}(2N)$ for model C. As a consequence of this 
correspondence, these reunion probabilities exhibit a third-order phase transition, akin to 
the Douglas-Kazakov transition in $YM_2$, as the size of the system $L$ crosses a critical 
value $L_c(N) \propto \sqrt{N}$ between the left tail $L < L_c(N)$ and the right tail $L > 
L_c(N)$. In the central part, $L \sim L_c(N)$, these reunion probabilities, converge in a 
proper scaling limit, when $N \to \infty$, to a limiting form which can be expressed in terms 
of the Tracy-Widom distributions ${\cal F}_1$ and ${\cal F}_2$: this fact was certainly one 
of the main results obtained in Ref. \cite{FMS11}. Here, we have provided a detailed and 
self-contained derivation of these results. The main emphasis of the paper is on the study of 
the large deviations of these reunion probabilities both in the right and in the left tail, 
together with a careful analysis of the matching between the different regimes (left tail, 
central part and right tail). While this matching in model A and model C is very similar to 
the one found for the distribution of the largest eigenvalue of GUE, one finds that the 
situation is much more involved in model B. In this case, there is instead a crossover (see 
Fig. \ref{fig_crossover}) between the central regime, for $L - 2\sqrt{N} \sim {\cal 
O}(N^{-1/6})$ and the right tail, $L > 2 \sqrt{N}$, of the reunion probability, this 
crossover happening at a crossover length scale $L_{\rm cross} - 2 \sqrt{N} \propto N^{-1/6} 
(\ln N)^{2/3}$, which seems to be a peculiar feature of these vicious walkers problems.

\begin{acknowledgement}
This research was partially supported by ANR grant 2011-BS04-013-01 WALKMAT and in part by the Indo-French 
Centre for the Promotion of Advanced Research under Project 4604-3. 
\end{acknowledgement}

\appendix

\section{Details about the large deviation regime}\label{app_large_dev}

In this appendix, we give some details concerning the calculation of $\tilde F_N(L)$ in the large deviation regime.

\subsection{Derivation of the formula given in Eq. (\ref{last_fn.1})}

We first provide a derivation of the formula given in Eq. (\ref{last_fn.1}) starting from (\ref{start.expr.2.1}) and (\ref{omega.orthogonal.2}). Indeed, from the definition of $R_k(\alpha)$ one has
\begin{eqnarray}
&&h_1(\alpha) = R_1(\alpha) h_0(\alpha) \;, \; h_2(\alpha) = R_2(\alpha) h_1(\alpha)= h_0(\alpha) R_1(\alpha) R_2(\alpha) \cdots 
\end{eqnarray}
and more generally
\begin{eqnarray}\label{expr_hk_app}
h_k(\alpha) = h_0(\alpha) R_1(\alpha) R_2(\alpha) \cdots R_k(\alpha) \;.
\end{eqnarray}
Therefore, $\Omega_N(\alpha)$ in Eq. (\ref{omega.orthogonal}) can be rewritten as
\begin{eqnarray}
\Omega(\alpha,N) &=& N! \prod_{j=1}^{N} h_{2j-1}(\alpha) = N! \, [h_0(\alpha) R_1(\alpha)] [h_0(\alpha) R_1(\alpha)R_2(\alpha)R_3(\alpha)] \nonumber \\
&\times& [h_0(\alpha) R_1(\alpha)R_2(\alpha)R_3(\alpha)R_4(\alpha)R_5(\alpha)] \cdots \nonumber \\
&=& N! [h_0(\alpha) R_1(\alpha)]^N [R_2(\alpha) R_3(\alpha)]^{N-1} [R_4(\alpha) R_5(\alpha)]^{N-2} \cdots \nonumber \\
&=&  N! [h_0(\alpha) R_1(\alpha)]^N \prod_{k=1}^{N-1} [R_{2k}(\alpha) R_{2k+1}(\alpha)]^{N-k}
\end{eqnarray}
which is the formula given in Eq. (\ref{last_fn.1}).

\subsection{Derivation of the formula given in Eq. (\ref{last_fn.2})}

In this appendix we give a detailed derivation, starting from the exact expression of $\tilde F_N(L)$ in Eq.~(\ref{start.expr.2.1}, \ref{start.expr.2.2}), of the asymptotic estimate for   
$\ln \tilde F_N(L)$, valid in the limit $L \gg \sqrt{2N} \gg 1$:
\begin{eqnarray}\label{last_fn.2_app}
\ln \tilde F_N(L) = e^{-\xi} \left[G_{2N}(\xi) + I_{2N}(\xi) \right] \;,
\end{eqnarray}  
in terms of the variable $\xi = \pi^2/\alpha = 2L^2$ and the functions $G_{2N}(\xi)$ given in Eq. (\ref{gk.laguerre}) and $I_{2N}(\xi)$ given by
\begin{eqnarray}\label{expr_I2N_app}
I_{2N}(\xi) = - 2 \xi \sum_{k=0}^{N-1} \frac{G_{2k+1}(\xi)}{2k+1} \;.
\end{eqnarray}
This formula (\ref{last_fn.2_app}) was given in Ref. \cite{CNS96} [see their Eq. (28)] without any detail: that is the purpose of this appendix to fill this gap by providing a detailed derivation of it.

To study the formula for $\tilde F_N(L)$ starting from Eq. (\ref{start.expr.2.1}, \ref{start.expr.2.2}), we first write the product of the gamma functions in the denominator as
\begin{eqnarray}\label{prod.gamma.1}
\prod_{j=0}^{N-1} \Gamma(2+j) \Gamma\left(\frac{3}{2}+j\right) = \prod_{k=1}^N k! \, \Gamma\left(\frac{1}{2}+k\right) \;.
\end{eqnarray} 
Note that the product of factorials can be written as
\begin{eqnarray}\label{prod.factorial}
\prod_{k=1}^N k! = 2^{N-1} 3^{N-2} \cdots N = N! \prod_{k=1}^{N-1} k^{N-k} \;. 
\end{eqnarray}
Similarly, the product of $\Gamma(k+1/2)$ in Eq. (\ref{prod.gamma.1}) can be written as
\begin{eqnarray}\label{prod.gamma.demi}
\prod_{k=1}^N \Gamma\left(k+\frac{1}{2} \right) = \left(\frac{\Gamma(1/2)}{2}\right)^N \prod_{k=1}^{N-1} (k+\frac{1}{2})^{N-k} \;,
\end{eqnarray}
where we have used 
\begin{eqnarray}
\Gamma(k+1/2) = \left(k-\frac{1}{2} \right) \left(k-\frac{3}{2} \right) \cdots \frac{1}{2} \Gamma\left(\frac{1}{2} \right) \;.
\end{eqnarray}
Finally, using Eqs (\ref{prod.factorial}) and (\ref{prod.gamma.demi}) we write Eq. (\ref{prod.gamma.1}) as
\begin{eqnarray}\label{prod.gamma.2}
\prod_{j=0}^{N-1} \Gamma(2+j) \Gamma\left(\frac{3}{2}+j\right)  = N! \left( \frac{\sqrt{\pi}}{2}\right)^N \prod_{k=1}^{N-1} k^{N-k} \prod_{k=1}^{N-1} (k+\frac{1}{2})^{N-k} \;,
\end{eqnarray}
where we have used $\Gamma(1/2) = \sqrt{\pi}$.

Using the ansatz for $R_k(\alpha)$ given in Eq. (\ref{rk.ansatz.largedev}) where we consider $c_k(\alpha) e^{-\pi^2/\alpha} \ll 1$, with $\alpha = \pi^2/2L^2$, we write, 
\begin{eqnarray} 
h_0(\alpha) R_1(\alpha) = \sqrt{\frac{\pi}{\alpha}} \frac{1}{2\alpha}\left(1+2e^{-\frac{\pi^2}{\alpha}} \right)\left(1+2 \alpha c_1(\alpha)e^{-\frac{\pi^2}{\alpha}} \right)
\end{eqnarray}
and for $k \geq 1$
\begin{eqnarray}
R_{2k}(\alpha) R_{2k+1}(\alpha) = \frac{k}{\alpha}\frac{k+1/2}{\alpha} \left[1 + \frac{\alpha}{k} c_{2k}(\alpha) e^{-\pi^2/\alpha}\right] \left[1 + \frac{\alpha}{k+1/2} c_{2k+1}(\alpha) e^{-\pi^2/\alpha}\right] + {\cal O}(e^{-2\pi^2/\alpha})  \;,
\end{eqnarray}
so that, using the formula in Eq. (\ref{prod.gamma.2}) one obtains that $\tilde F_N(L)$ in Eq. (\ref{start.expr.2.1}, \ref{start.expr.2.2}) can be written as
\begin{eqnarray}\label{fn.app.large.dev}
\tilde F_N(L) &=& \left( 1 + 2e^{-\frac{\pi^2}{\alpha}} \right)^N \left(1 + 2 \alpha c_1(\alpha) e^{-\frac{\pi^2}{\alpha}}\right)^N \\
&\times& \prod_{k=1}^{N-1} \left( 1 + \frac{\alpha}{k} c_{2k}(\alpha) e^{-\frac{\pi^2}{\alpha}} \right)^{N-k}  \prod_{k=1}^{N-1} \left( 1 + \frac{\alpha}{k+1/2} c_{2k+1}(\alpha) e^{-\frac{\pi^2}{\alpha}} \right)^{N-k} \;. \nonumber
\end{eqnarray}
We now perform the expansion of this expression (\ref{fn.app.large.dev}), considering $c_k(\alpha) e^{-\pi^2/\alpha} \ll 1$. It is more convenient to expand its logarithm $\ln \tilde F_N(L)$ which using the explicit expression for $c_1(\alpha)$ in Eq. (\ref{c01}) together with the expression of $c_k(\alpha)$ in terms of $G_k(\xi)$ in Eq. (\ref{eq:ck_gk}), can be written as a function of $\xi = \pi^2/\alpha$ as
\begin{eqnarray}\label{fn.app.large.dev.2}
\ln \tilde F_N(L) &=& - 2N(2 \xi-1)e^{-\xi} - 2 \xi e^{-\xi} \sum_{k=1}^{N-1} \left( \frac{N-k}{k} G_{2k}(\xi) + \frac{N-k}{k+1/2} G_{2k+1}(\xi)  \right) \nonumber \\
&=& 2 Ne^{-\xi} - 2 \xi e^{-\xi} \left( \sum_{k=1}^{N-1} \frac{2N-2k}{2k} G_{2k}(\xi) + \sum_{k=0}^{N-1} \frac{2N-2k}{2k+1} G_{2k+1}(\xi)  \right)
\end{eqnarray} 
where we have used $G_1(\xi) = 1$. It is then possible to write these two sums 
in Eq. (\ref{fn.app.large.dev.2}) as
\begin{eqnarray}\label{fn.app.large.dev.3}
\ln \tilde F_N(L) &=& 2N\, e^{-\xi} - 2 \xi e^{-\xi} \left( \sum_{j=1}^{2N-1}  
\frac{2N-j}{j} G_{j}(\xi) + \sum_{k=0}^{N-1} \frac{G_{2k+1}(\xi)}{2k+1} \right) \;.
\end{eqnarray}
Finally, using the following identity satisfied by the polynomials $G_k(\xi)$~\cite{GM94} 
\begin{eqnarray}
2N -2 \xi \sum_{j=1}^{2N-1} \frac{2N-j}{j} G_j(\xi) = G_{2N}(\xi) \;,
\end{eqnarray}
which can be shown, for instance, by using their explicit expression of $G_k(\xi)$ (\ref{gk.laguerre}), one obtains finally
\begin{eqnarray}
\ln \tilde F_N(L) = e^{-\xi} \left(G_{2N}(\xi) + I_{2N}(\xi) \right) \;,
\end{eqnarray}
as given in the text in Eq. (\ref{last_fn.2}).

\subsection{An integral representation for $I_{2N}(\xi)$ in Eq. (\ref{expr_I2N})}

We start with the integral representation for $G_k(\xi)$ in Eq. (\ref{contour.gk}) to express $I_{2N}(\xi)$ in Eq. (\ref{expr_I2N}) as
\begin{eqnarray}\label{I2N.app.1}
I_{2N}(\xi) = -2\xi\oint_{C_t} \frac{dt}{2\pi i} e^{-2 \xi t} \sum_{k=0}^{N-1} \frac{1}{2k+1} \left(1 + \frac{1}{t} \right)^{2k+1} \;.
\end{eqnarray}
Performing an integration by part one obtains
\begin{eqnarray}\label{I2N.app.2}
I_{2N}(\xi) = \oint_{C_t} \frac{dt}{2\pi i} \frac{e^{-2 \xi t}}{t^2} \sum_{k=0}^{N-1} \left(1 +\frac{1}{t} \right)^{2k} \;.
\end{eqnarray}

By performing the sum over $k$ in (\ref{I2N.app.2}) and dropping an $N$-independent constant
term, 
one finally arrives at the expression 
given in the text in Eq.~(\ref{i_integral}).

\end{document}